\documentclass[reprint,aps,prl,amsmath,amssymb,twocolumn,showkeys, superscriptaddress,floatfix,showkeys]{revtex4-2}
\usepackage{graphicx}
\usepackage{dcolumn}
\usepackage{bm}
\usepackage[colorlinks = true, allcolors = blue]{hyperref}
\usepackage{verbatim}
\usepackage{soul}
\usepackage{float}
\usepackage{xcolor}
\usepackage{multirow}
\usepackage{tikz}
\usetikzlibrary{shapes.geometric, arrows, shadows,positioning}

\tikzstyle{startstop} = [rectangle, rounded corners, 
minimum width=3cm, 
minimum height=0.75cm,
text centered, 
draw=black, 
fill=red!30]

\tikzstyle{io} = [trapezium, 
trapezium stretches=true, 
trapezium left angle=70, 
trapezium right angle=110, 
minimum width=3cm, 
minimum height=0.75cm, text centered, 
draw=black, fill=blue!30]

\tikzstyle{process} = [rectangle, 
minimum width=3cm, 
minimum height=0.75cm, 
text centered, 
draw=black, 
fill=orange!30]

\tikzstyle{decision} = [diamond, 
minimum width=3cm, 
minimum height=0.75cm, 
text centered, 
draw=black, 
fill=green!30]
\tikzstyle{arrow} = [thick,->,>=stealth]

\begin{document}
\title{Nonlinear effects in many-body van der Waals interactions}

\author{Dai-Nam Le}
\email{dainamle@usf.edu}
\affiliation{Department of Physics, University of South Florida, Tampa, Florida 33620, USA}

\author{Pablo Rodriguez-Lopez}
\affiliation{{\'A}rea de Electromagnetismo and Grupo Interdisciplinar de Sistemas Complejos (GISC), Universidad Rey Juan Carlos, 28933, M{\'o}stoles, Madrid, Spain}
\affiliation{Laboratoire Charles Coulomb (L2C), UMR 5221 CNRS-University of Montpellier, F-34095 Montpellier, France}

\author{Lilia M. Woods} 
\email{lmwoods@usf.edu}
\thanks{Author to whom all correspondence should be addressed}
\affiliation{Department of Physics, University of South Florida, Tampa, Florida 33620, USA}

\date{\today}

\begin{abstract}
Van der Waals interactions are ubiquitous and they play an important role for the stability of materials. Current understanding of this type of coupling is based on linear response theory, while optical nonlinearities are rarely considered in this context. Many materials, however, exhibit strong optical nonlinear response, which prompts further evaluation of dispersive forces beyond linear response. Here we present a \textit{Discrete Coupled Nonlinear Dipole} approach that takes into account linear and nonlinear properties of all dipolar nanoparticles in a given system. This method is based on a Hamiltonian for nonlinear dipoles, which we apply in different systems uncovering a complex interplay of distance, anisotropy, polarizibilities, and hyperpolarizabilities in the vdW energy. This investigation broadens our basic understanding of dispersive interactions, especially in the context of nonlinear materials. 
\end{abstract}
\maketitle

\paragraph{Introduction ---}Van der Waals (vdW) interactions play an essential role for the stability of materials with chemically inert components \cite{Gjerding_2021,Geng2018}. The vdW interaction is important for the stacking patterns of layered materials \cite{Pandey2020,Barik2022}, and  it can even lead to significant changes in the electronic structure as is the case for the direct-to-indirect band gap transition when comparing a monolayer MoS$_2$ and its bulk counterpart \cite{Yu2018}. First principles computational methods for calculating vdW interactions within linear response theory have also been developed \cite{Ambrosetti2016, Maurer2019} with implementations in codes, such as VASP \cite{Kresse1996, Kresse1996b} and Quantum Espresso \cite{Giannozzi2017}. In these packages, however, vdW calculations rely on linear response theory, which may be problematic for materials with optical nonlinearities. 

Recently, several noncentrosymmetric transition metal dichalcogenides \cite{Li2013, Wang2015, Autere2018, Wen2019} or Weyl semimetals \cite{Patankar2018, Takasan2021, Drueke2021, Xu2020} have been found to have much enhanced second-order nonlinear hyperpolarizability, which is important in novel applications for the coherent control of spin and valley polarized currents. Many of these materials can also exhibit large third-order optical nonlinearities, which is beneficial for ultrafast optics and plasmonics \cite{You2019,Zhou2020}. Given that optical properties play an essential role in the collective nature of vdW interactions, one expects that hyperpolarizbilities will also affect the coupling between materials. Indeed, recent electrodynamical studies have shown that even at large separations the interaction between macroscopic bodies experiences magnitude and/or scaling law modulations depending on their nonlinear properties \cite{Soo2016, Soo2018,Soo2017}. VdW quadrupolar contributions have also been examined with the random phase approximation showing that these beyond-dipolar effects cannot be neglected in molecular systems \cite{Massa2021}. 

Main-stream computational schemes currently do not take into account optical nonlinearities \cite{Kresse1996, Kresse1996b,Giannozzi2017}, which may be a hurdle for a more accurate calculations for electronic structure properties and phenomena directly related to vdW interactions. It is known that even within linear response theory, the vdW coupling energy can have vastly different magnitude, scaling law, and/or sign than the typical attractive two-body London-type $1/R^6$ behavior \cite{Stedman2014,Hopkins2014,Cole2009}. A striking example are carbon materials, where the vdW forces can be quite different for carbon nanotubes, graphene layers, or graphene nanoribbons \cite{Tkatchenko2013b,Sarabadani2011,Drosdoff2014}. This is a consequence of the complex interplay between many factors: separation, atomic distributions, optical response, and many-body effects. 

Here we present the \textit{Discrete Coupled Nonlinear Dipole} method as a generalized framework for computing dispersive interactions between nanoparticles. In this many-body approach the linear and nonlinear optical properties for all interacting nanoparticles are taken into account explicitly. For this purpose, we obtain the quantum mechanical Hamiltonian $\hat{H}$ for interacting \textit{nonlinear dipoles} fluctuating around their equilibrium positions, which was previously unavailble. Our scheme relies on a diagnonalization procedure coupled with perturbation theory to obtain the eigenmodes of $\hat{H}$ whose sum is the vdW interaction energy. The method allows a miscroscopic dissection of dispersive coupling in terms of anisotropy, tensorial nature of the optical response, inversion symmetry, and distance separations for interacting materials.

\paragraph{Theoretical Model ---} 
We consider a system composed of $N$ nanoparticles with no external fields applied. Each nanoparticle has a dipolar moment $\hat{\mathbf{p}}^{i}$ ($i=1,2,...,N$) generated by the induced electric field from the other nanoparticles. The Hamiltonian of this system is $\hat{H} = \sum\limits_{i = 1}^N \hat{H}^{i} - \frac{1}{2} \sum\limits_{i,j=1}^{N} \hat{\mathbf{p}}^{i} \mathbf{T}^{ij} \hat{\mathbf{p}}^{j} $, where the displacement matrix $\mathbf{T}^{i j} = \frac{1}{4\pi \varepsilon_0} \frac{3 \mathbf{R}^{i j} \mathbf{R}^{i j} - (R^{i j})^2 \mathbf{1}}{(R^{i j})^5}$ is determined by the separation vector $\mathbf{R}^{i j}=\mathbf{R}^{i}-\mathbf{R}^{j}$. 

Since previously the quantum mechanical Hamiltonian for an individual nanoparticle $\hat{H}^{j}$ is available only from linear response, we begin with the derivation of $\hat{H}^{j}$ for a single nonlinear dipole. The starting point is a classical dipole for particle $i$ in an electric field $\mathbf{E}$: $\mathbf{p}^{i} = \pmb{\alpha}^{i} \mathbf{E}  + \pmb{\beta}^{i} \mathbf{E} \mathbf{E} +  \pmb{\gamma}^{i}  \mathbf{E} \mathbf{E} \mathbf{E} +...$, where $\pmb{\alpha}^{i}$ is the polarizability tensor of rank 2 from linear response, while $\pmb{\beta}^{i}$, $\pmb{\gamma}^{i},...$ are second-order, third-order,... hyperpolarizabilities (rank 3, 4,... tensors) \cite{Lozovskii1990, Lozovskii1990b, Makhnovets2016,Boyd2007}. From the classical Hamilton and the equations of motion for $\mathbf{p}^{i}$ and its canonical momentum $\mathbf{\pi}^{i}$ taken under equilibrium conditions, a self-consistent nonlinear equation for $\mathbf{p}^{i}$ is found. Further, assuming small dipole moments oscillating with a characteristic frequency $\omega^0$, the solutions to the Hamilton equations combined with a standard quantization procedure are used to transform the classical Hamiltonian into its quantum mechanical equivalent,

\begin{widetext}
\begin{eqnarray}
\label{eqn:full-hamiltionian-matrix-0}
\hat{H} && = \frac{1}{2} \mathbb{F}_{mn} \hat{\Pi}_m \hat{\Pi}_n + \frac{1}{2} \mathbb{A}_{mn} \hat{P}_m \hat{P}_n + \frac{1}{3} \mathbb{B}_{mnq} \hat{P}_m \hat{P}_n \hat{P}_q + \frac{1}{4} \mathbb{G}_{mnqp} \hat{P}_m \hat{P}_n \hat{P}_q \hat{P}_p , \\
\label{eqn:F-tensor}
\mathbb{F}_{mn} && = \left[(\omega^{0})^2 \pmb{\alpha} \right] \delta_{ \left\lfloor \frac{m+2}{3} \right\rfloor \left\lfloor \frac{n+2}{3} \right\rfloor  } , \quad \mathbb{A}_{mn} = \left[ \pmb{\alpha}^{-1} \right] \delta_{ \left\lfloor \frac{m+2}{3} \right\rfloor \left\lfloor \frac{n+2}{3} \right\rfloor  } - \mathbf{T}^{ \left\lfloor \frac{m+2}{3} \right\rfloor \; \left\lfloor \frac{n+2}{3} \right\rfloor }  \left(1 - \delta_{ \left\lfloor \frac{m+2}{3} \right\rfloor \left\lfloor \frac{n+2}{3} \right\rfloor  } \right), \\ 
\label{eqn:B-tensor}
\mathbb{B}_{mnq} && = - \left[ \pmb{\alpha}^{-1} \otimes \pmb{\beta} \otimes \pmb{\alpha}^{-1} \otimes \pmb{\alpha}^{-1} \right] \delta_{ \left\lfloor \frac{m+2}{3} \right\rfloor \left\lfloor \frac{n+2}{3} \right\rfloor \left\lfloor \frac{q+2}{3} \right\rfloor } ,\\
\label{eqn:G-tensor}
\mathbb{G}_{mnqp} && = \left[ 2 \pmb{\alpha}^{-1} \otimes \pmb{\beta} \otimes \pmb{\alpha}^{-1} \otimes \pmb{\alpha}^{-1} \otimes \pmb{\beta} \otimes \pmb{\alpha}^{-1} \otimes \pmb{\alpha}^{-1} - \pmb{\alpha}^{-1}\otimes \pmb{\gamma} \otimes \pmb{\alpha}^{-1} \otimes \pmb{\alpha}^{-1} \otimes \pmb{\alpha}^{-1}\right] \delta_{ \left\lfloor \frac{m+2}{3} \right\rfloor \left\lfloor \frac{n+2}{3} \right\rfloor \left\lfloor \frac{q+2}{3} \right\rfloor \left\lfloor \frac{p+2}{3} \right\rfloor },
\end{eqnarray}
\end{widetext}
where $m,n,q,p = 1, 2, \ldots, 3N$ denote the degrees of freedom associated with $x, y ,z$ directional property components of the $N$ nanoparticles. Also, $\left\lfloor \frac{m+2}{3} \right\rfloor$ being the integer part of $\frac{m+2}{3}$ tracks the nanoparticle and its $m^{th}$ dipolar component. A generalized Kronecker delta notation $\delta_{ a b c d ... } = \left\{ \begin{array}{ll}
1 & \text{ when } a = b = c = d = ... \\
0 & \text{ otherwise}
\end{array} \right.$ and the Einstein summation rule are also used in Eqs. \eqref{eqn:full-hamiltionian-matrix-0}--\eqref{eqn:G-tensor}.  A detailed derivation of $\hat{H}^j$ and $\hat{H}$ is given in the Supplementary Information \cite{supplementary}. The first term in Eq. \eqref{eqn:full-hamiltionian-matrix-0} contains the kinetic energy of the dipoles described by a $1\times 3N$ column matrix for the canonical momenta $\hat{\Pi}$ with components $\hat{\Pi}_m$. The second term corresponds to the linear response properties of the fluctuating dipolar moments arranged  in a $1\times 3N$ column matrix $\hat{P}$ with components $\hat{P}_m$ \cite{Balla2010, Balla2012}. The third term contains the $\pmb{\beta}$ properties only, while the last term includes both $\pmb{\beta}$ and $\pmb{\gamma}$ nonlinearities. 

\begin{figure}[htbp]
\centering
\begin{tikzpicture}[auto, node distance=1.3cm]

\node (in1) [io, align = center] {Input parameters $ \pmb{\alpha}, \pmb{\beta}, \pmb{\gamma} \rightarrow \mathbb{F}, \mathbb{A}, \mathbb{B}, \mathbb{G}$ \\
Minimize $\frac{\partial \hat{H}}{\partial \hat{P}_m} (P_{0,m}) = 0 \rightarrow P_{0,m}, E_0 = H(P_{0,m})$\\
$\hat{H} = E_0 + \hat{H}_{h} + \hat{H}_{anh}$
};

\node (pro2) [process, below of=in1, align = center, yshift = -0.3 cm] {Solve Discrete Coupled Dipole equation\\
$\left[ \mathbb{A}_{mn} + 2 \mathbb{B}_{mnq} P_{0,q} + 3 \mathbb{G}_{mnqp} P_{0,q} P_{0,p} \right]\mathbb{F}_{nl} \mathbf{u}_l = \omega ^2 \mathbf{u}_l$\\
$\mathbb{F}_{m n} u_{m q} u_{n p} = \delta_{q p}$
};

\coordinate[below of=pro2] (Empty1);

\node (pro3) [process, left of=Empty1, align = center, xshift = -0.5 cm] {Eigenvalues $\omega_{n}$\\
$E_{g.s}^{(0)} = \sum_{n=1}^{3N} \frac{1}{2} \hbar \omega_n$
};

\node (pro4) [process, right of=Empty1, align = center, xshift = 0.5 cm] {Eigenvectors $\mathbf{u}_n$\\
$\mathbb{M} = \mathbb{F} \left[ \mathbf{u}_1 \; \mathbf{u}_2 ... \mathbf{u}_{3N} \right]$
};

\coordinate[below of=Empty1, yshift = 0.1 cm] (Empty2);

\node (pro5) [process, below of=pro4, align = left, yshift = 0.1 cm] {
Evaluate $\Delta E_{2nd}^{(2)}$, $\Delta E_{3rd}^{(1)}$\\
with perturbation theory
};

\coordinate[below of=Empty2] (Empty3);

\node (out1) [io, left of=Empty3, align = center, xshift = -0.5 cm] {
$E_{g.s} = E_0 + E_{g.s}^{(0)}$\\
$+ \Delta E_{2nd}^{(2)} + \Delta E_{3rd}^{(1)}$
};
\node (stop) [startstop, right of=Empty3, align = center, xshift = 0.5 cm] {
$U_{vdW} = E_{A+B}$\\
$- E_{A} - E_{B}$};

\draw [arrow] (in1) -- (pro2);

\draw [arrow] (pro2) -- (pro3);

\draw [arrow] (pro2) -- (pro4);

\draw [arrow] (pro4) -- (pro5);

\draw [arrow] (pro3) -- (out1);

\draw [arrow] (pro5) -- (out1);

\draw [arrow] (out1) -- (stop);

\end{tikzpicture}
\caption{\label{fig:flowchart} Schematic flowchart of the \textit{Discrete Coupled Nonlinear Dipole} method.}
\end{figure}
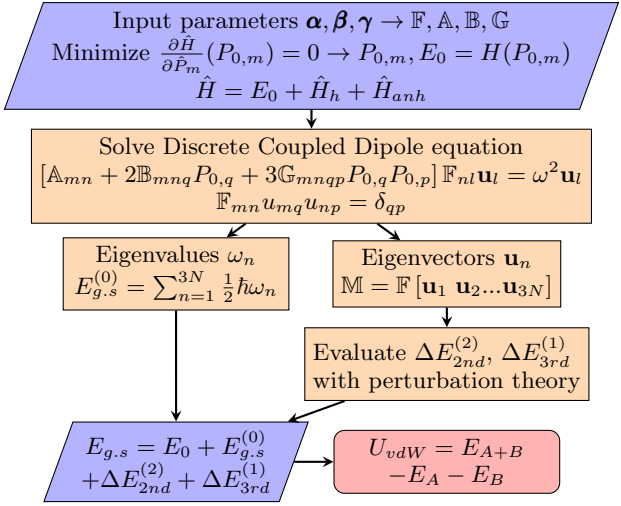

The Hamiltonian in Eq. \eqref{eqn:full-hamiltionian-matrix-0} sets the stage for the vdW energy calculations that takes into account the linear and nonlinear response of each nanoparticle. The schematic flowchart in Fig. \ref{fig:flowchart} gives the main steps of the method discussed in what follows. After identifying the optical properties of each nanoparticle, we represent $\hat{P}_m=P_{0,m} + \hat{Q}_m$, where the equilibrium position $P_{0,m}$ is found from $\dfrac{\partial \hat{H}}{\partial \hat{P}_m} = 0$ and 
$\hat{Q}_m$ is the fluctuation of each dipole around $P_{0,m}$. This enables transforming the Hamiltonian into $\hat{H} = E_0 + \hat{H}_{h} + \hat{H}_{anh}$ where $E_0=\hat{H}(P_{0,m})$ is the minimum free energy at equilibrium, $\hat{H}_{h}$ contains $\hat{\Pi}_m \hat{\Pi}_n$ and $\hat{Q}_m \hat{Q}_n$ terms, while $\hat{H}_{anh}$ consists of $\hat{Q}_n \hat{Q}_m \hat{Q}_q$ and $\hat{Q}_n \hat{Q}_m \hat{Q}_q \hat{Q}_p$ terms \footnote{$E_0 = \frac{1}{2} \mathbb{A}_{mn} {P}_{0,m} {P}_{0,n} + \frac{1}{3} \mathbb{B}_{mnq} {P}_{0,m} {P}_{0,n} {P}_{0,q} + \frac{1}{4} \mathbb{G}_{mnqp} {P}_{0,m} {P}_{0,n} {P}_{0,q} {P}_{0,p}$, $\hat{H}_{h} = \frac{1}{2} \left[ \mathbb{A}_{mn} + 2 \mathbb{B}_{mnq} P_{0,q} + 3 \mathbb{G}_{mnqp} P_{0,q} P_{0,p} \right] \hat{Q}_m \hat{Q}_n$ and $\hat{H}_{anh}^{\prime} = \frac{1}{3} \left( \mathbb{B}_{mnq} + 3 \mathbb{G}_{mnqp} P_{0,p} \right) \hat{Q}_{m} \hat{Q}_{n} \hat{Q}_{q} + \frac{1}{4} \mathbb{G}_{mnqp} \hat{Q}_{m} \hat{Q}_{n} \hat{Q}_{q} \hat{Q}_{p}$ is the anharmonic Hamiltonian}. From the diagonalization of $\hat{H}_{h}$, we find a system of coupled equations for the dipolar fluctuations,
\begin{eqnarray}
   \label{eqn:eigenvalue-explicit-full}
   \left[ \mathbb{A}_{mn} + 2 \mathbb{B}_{mnq} P_{0,q} + 3 \mathbb{G}_{mnqp} P_{0,q} P_{0,p} \right] \mathbb{F}_{nl} \mathbf{u}_l = \omega ^2 \mathbf{u}_l
\end{eqnarray}
where the eigenvectors $\mathbf{u}_l$ satisfy the normalization condition $\mathbb{F}_{mn} u_{mq} u_{np} = \delta_{qp}$. The eigenvalues from Eq. \eqref{eqn:eigenvalue-explicit-full} 
constitute the zero-point modes which give the zeroth order ground state energy $E_{g.s}^{(0)} = \sum\limits_{n=1}^{3N} \frac{1}{2} \hbar \omega_n$. Further, considering that linear response is always stronger than the optical nonlinearities, $\hat{H}_{anh}$ is treated as a perturbation finding that 
\begin{eqnarray}
    \label{eqn:groundstate-energy-full}
    E_{g.s} \approx E_0 + E_{g.s}^{(0)} + \Delta E_{2nd}^{(2)} + \Delta E_{3rd}^{(1)}, 
\end{eqnarray}
where $\Delta E_{2nd}^{(2)}$ is the second-order correction coming from $\hat{Q}_n \hat{Q}_m \hat{Q}_q$ terms in $\hat{H}_{anh}$ (the first order correction is zero) and $\Delta E_{3rd}^{(1)}$ is the first-order correction from $\hat{Q}_n \hat{Q}_m \hat{Q}_q \hat{Q}_p$ terms in $\hat{H}_{anh}$ (details in the Supplementary Information \cite{supplementary}).

The approach in Fig. \ref{fig:flowchart} constitutes the {\it Discrete Coupled Nonlinear Dipole} method, which can now be applied to any system composed of discrete nonlinear dipoles. Note that when $\mathbb{B}_{mnq}=\mathbb{G}_{mnqp}=0$, one recovers the linear Discrete Coupled Dipole method \cite{Shtogun2010}, which was already successfully used  in first principles many-body schemes \cite{Silvestrelli2008, Tkatchenko2013, Tkatchenko2014, Tkatchenko2019, Tkatchenko2022} for calculating vdW energy within {\it linear} dipolar interactions. The generalized method described here provides means to account explicitly for {\it linear and nonlinear} properties in many-body vdW interactions.

\begin{figure}[htbp]
    \centering
    \includegraphics[width = \columnwidth]{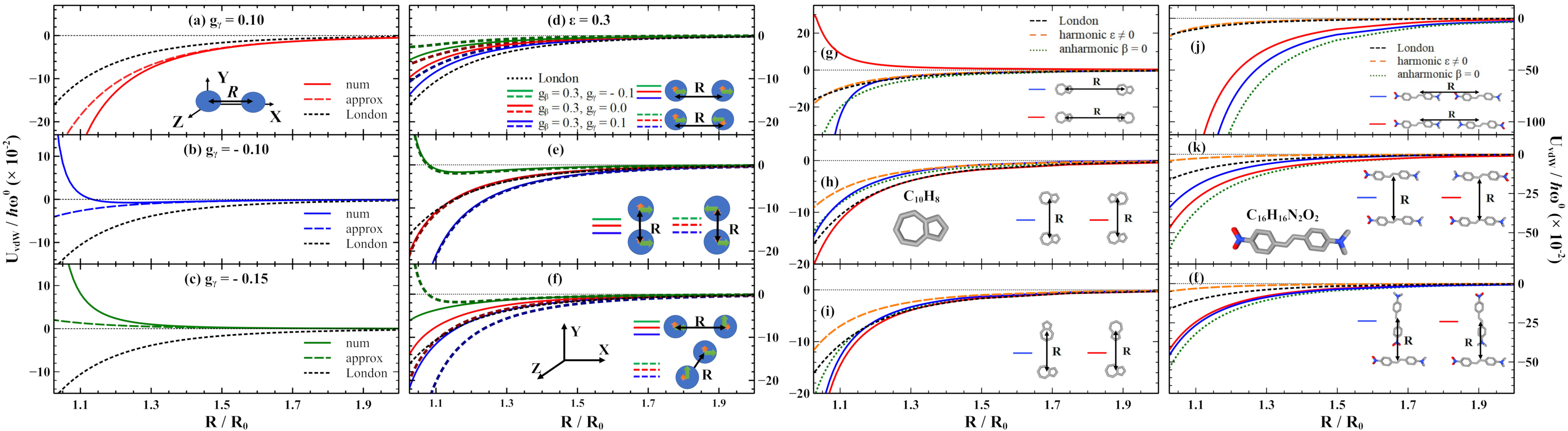}
    \caption{\label{fig:2}  VdW energy rescaled by $\hbar \omega^0$ vs nanoparticle separation rescaled by $R_0=(\alpha/2\pi\epsilon_0)^{1/3}$ is shown in all panels. Two isotropic nanoparticles with (a) $g_{\gamma} =0.1$; (b) $g_{\gamma} =-0.1$; (c) $g_{\gamma} =-0.15$ with numerical (num), Eq. \eqref{eqn:vdW-case1b-approx} (approx), and London formula $U_L = -(3/16)\hbar \omega^0 (R_0/R)^6$ results. Two anisotropic nanoparticles with $\epsilon = 0.3$, $g_\beta=0.3$ and $g_\gamma=0, \pm 0.1$ with a relative angle between their optical axis (d) $\psi=0$ (solid), $\psi=\pi$ (dashed) and separation along $x$-axis; (e) $\psi=0$ (solid), $\psi=\pi$ (dashed) and separation  along $y$-axis; (f) $\psi=\frac{\pi}{2}$ with separation  along $x$-axis (solid) and $\psi=\frac{\pi}{2}$ with separation along $z$-axis (dashed). Two azulenes (C$_{10}$H$_{8}$) with (g) $\psi=0$ (blue) and $\psi=\pi$ (red) and separation along $x$-axis; (h) $\psi=0$ (blue) and $\psi=\pi$ (red) and separation along $y$-axis; (i) $\psi=\frac{\pi}{2}$ (blue) and $\psi=-\frac{\pi}{2}$ (red) and separation along $y$-axis. Two 4-dimethylamino-4'-nitrostilbenes (C$_{16}$H$_{16}$N$_2$O$_2$) with (g) $\psi=0$ (blue) and $\psi=\pi$ (red) and separation along $x$-axis; (h) $\psi=0$ (blue) and $\psi=\pi$ (red) and separation along $y$-axis; (i) $\psi=\frac{\pi}{2}$ (blue) and $\psi=-\frac{\pi}{2}$ (red) and separation along $y$-axis. Numerical calculations with all molecular linear and nonlinear properties (full lines), neglecting all nonlinear properties (harmonic $\epsilon\neq 0$), neglecting only second hyperpolarizability (anharmonic $\beta=0$) are shown.  Here, $g_{\gamma} = \gamma \hbar \omega^0 / \alpha^2$ and $g_\beta= \beta(\hbar \omega^0)^{1/2} / (\alpha (1-\epsilon))^{3/2}$.}
\end{figure}

\paragraph{Isotropic particles ---} The method is now applied to two identical isotropic nanoparticles placed at $(0,0,0)$ and $(R,0,0)$ (insert in Fig. \ref{fig:2}a) whose linear polarizabilities are  $\pmb{\alpha}^{1}_{\mu \nu} = \pmb{\alpha}^{2}_{\mu \nu} = \alpha \delta_{\mu \nu} $. For the sake of simplicity, we further take that $\pmb{\gamma}^{1}_{\mu \nu \lambda \rho} = \pmb{\gamma}^{2}_{\mu \nu \lambda \rho} = \frac{\gamma}{3}  \left( \delta_{\mu \nu} \delta_{\lambda \rho} +  \delta_{\mu \lambda} \delta_{\nu \rho} +  \delta_{\mu \rho} \delta_{\nu \lambda} \right)$. Note that in this case $\pmb{\beta}^{1} = \pmb{\beta}^{2}=0$ due to the presence of inversion symmetry in isotropic materials \cite{Boyd2007}. Results from the numerical calculations for distances larger than the size of the dipole $R_0 = \sqrt[3]{\frac{\alpha}{2\pi\varepsilon_0}}$ are shown in Fig. \ref{fig:2}a, where  $\quad g_{\gamma} =  \frac{\gamma \hbar \omega^{0}}{ \alpha^2} $ "measures" the relative nonlinear-to-linear response strength  (for materials, $|g_{\gamma}|<1$ \cite{Boyd2007}). We find that $g_{\gamma}$ can result in significant modifications of the vdW energy, which are much pronounced at small separations. In the case of $\gamma>0$, there is a stronger vdW attraction (Fig. \ref{fig:2}a), but for $\gamma<0$, the vdW interaction is repulsive (Fig. \ref{fig:2}b,c). For $R \gg R_0$ we are able to obtain an analytical expression (details in the Supplementary Information \cite{supplementary}) 
\begin{eqnarray}
    \label{eqn:vdW-case1b-approx}
    U_{vdW} (R) \approx \left(1 + \frac{15}{2} g_{\gamma} \right) U_{L} (R),
\end{eqnarray}
where $U_{L} (R)=- \frac{3}{16} \hbar \omega^0 \frac{R_0^6}{R^6}$ is the standard London result for two atoms \cite{London1937}. Eq. \eqref{eqn:vdW-case1b-approx}, which agrees very well with the numerical calculations for $\sim\frac{R}{R_0}>1.5$ as given in Fig. \ref{fig:2}a-c, shows that the nonlinear response does not change the $1/R^6$ London-like behavior. However, $g_\gamma$ modules the strength of the interaction and it can even result in repulsion when $g_{\gamma}<-2/15$.

\onecolumngrid

\twocolumngrid

\paragraph{Anisotropic particles ---} We also consider anisotropic nanoparticles taken to have two distinct optical axes with $\alpha_{yy}=\alpha_{zz}=\alpha_{\parallel}=\alpha$ and $\alpha_{xx}=\alpha_{\perp} = \alpha(1-\epsilon)$ where $0 \leq \epsilon \leq 1$. Many anisotropic materials, such as  noncentrosymmetric transition metal dichalcogenides, polar metals, and Weyl semimetals \cite{Li2013, Wang2015, Autere2018, Wen2019, Patankar2018, Takasan2021, Drueke2021, Xu2020}, have broken inversion symmetry and they exhibit strong second and third order  nonlinear response.  Although much interest has been generated due to their potential applications for optical rectification and second harmonic generation applications, for example, here we show that the effect of nonlinearity is ``felt'' at a fundamental level concerning dispersive interactions. 

To illustrate these nonlinear effects in the vdW energy, we first focus on two identical nanoparticles with  $\pmb{\beta}_{xxx} = \beta$ (all other $\pmb{\beta}$ components are zero) and  $\pmb{\gamma}_{yyyy} = \pmb{\gamma}_{zzzz} = 3 \pmb{\gamma}_{yyzz} = \gamma$ (all other $\pmb{\gamma}$ components are zero). In addition to $R_0$ and $g_\gamma$, we introduce $g_{\beta} = \frac{ \beta \left( \hbar \omega^0 \right)^{1/2} }{\left( \alpha (1 - \epsilon) \right)^{3/2}}$. The vdW dependence on the optical axis relative orientation is tracked by the angle $\psi$ (inserts in Fig. \ref{fig:2}d-f ). The results in Fig. \ref{fig:2}d-f show that the vdW energy depends collectively on the anisotropy (parameter $\epsilon$), the optical nonlinearities ($g_\beta$ and $g_\gamma$ parameters), and the optical axis orientation (angle $\psi$). For particles separated along the $x$-axis, the nonlinear response leads to a reduced attraction as compared to the London formula. This reduction is strongest for $g_\beta$ and $g_\gamma$ with opposite signs (Fig. \ref{fig:2}d). For particles separated along the $y$-axis and $\psi=\frac{\pi}{2}$, $g_\beta$ and $g_\gamma$ lead to stronger vdW attraction, while particles on the $z$-axis and $\psi=\frac{\pi}{2}$ experience reduced attraction when compared with $U_L$ (Fig. \ref{fig:2}f). It is also possible to obtain a repulsive vdW energy when the particles are on the $y$-axis but their optical axis have a relative enagle $\psi=\pi$
(Fig. \ref{fig:2}e). The numerical results given in Fig. \ref{fig:2}d-f are consistent with the analytical expressions found in the limit of $R \gg R_0$ (see in Section S-V and Fig. S2 on  Supplementary Information \cite{supplementary} for detailed derivations and comparisons),

\begin{eqnarray}\label{eqn:case2-asymptotic}
&&  U_{vdW} \left( \mathbf{R} \right) \approx \nonumber\\
&& \approx \left\{
\begin{array}{l}
\left( 1 - \frac{2 \epsilon (2 - \epsilon)}{3} + 2 g_{\gamma} - \frac{653 g_{\beta}^2 (1-2\epsilon) }{162} \right) U_L (R) \\
\quad  \mp  \frac{g_{\beta}^2\hbar \omega^0}{16} \frac{R_0^3}{R^3}, \quad \quad \mathbf{R} = R \mathbf{e}_x, \psi = 0(-), \pi(+) \\
\left( 1 - \frac{\epsilon(2 - \epsilon)}{6} + 5 g_{\gamma} - \frac{653 g_{\beta}^2 (1-2\epsilon)}{648}  \right) U_L (R) \\
 \quad \pm \frac{g_{\beta}^2 \hbar \omega^0}{32} \frac{R_0^3}{R^3}, \quad \quad  \mathbf{R} = R \mathbf{e}_y, \psi = 0(+), \pi(-) \\
\left( 1 - \frac{5 \epsilon}{6} + \frac{(7 - 5 \epsilon)g_{\gamma}}{2} - \frac{5153 g_{\beta}^2 (1-\epsilon) }{2592} \right) U_L (R) \\
 \quad \quad \quad \quad\quad \quad\quad \quad \mathbf{R} = R \mathbf{e}_x, \psi = \frac{\pi}{2}\\
\left( 1 - \frac{\epsilon}{3} + (5 -  \epsilon) g_{\gamma} - \frac{653 g_{\beta}^2 (1-\epsilon) }{648} \right) U_L (R) \\
 \quad \quad \quad \quad\quad \quad\quad \quad \mathbf{R} = R \mathbf{e}_y, \psi = \frac{\pi}{2}
\end{array}   
       \right..
\end{eqnarray} 
The above expression reflects the complexity of the vdW interaction for anisotropic particles. There is always a London-like term $U_L(R)$, but rescaled by a factor that depends nontrivially  on $\epsilon$, $g_{\beta}$, $g_\gamma$, and $\psi$. The vdW energy also has a term entirely due to the second order nonlinearity, which has a much longer outreach range due its $1/R^3$ dependence. This term can be positive or negative (dictated mainly by $\psi$) and it can be a decisive factor for the energy behavior especially at larger $R$. Eq. \eqref{eqn:case2-asymptotic} further shows that by tailoring the properties of the nanoparticles, the London contribution can be suppressed in which case the interaction is dominated by the long-ranged $1/R^3$ dependence as determined by the second order nonlinearity. 

The \textit{Discrete Coupled Nonlinear Dipole} model is also applied to realistic systems taking azulene (C$_{10}$H$_{8}$) and 4-dimethylamino-4'-nitrostilbene (C$_{16}$H$_{16}$N$_2$O$_2$) as representatives for nonlinear dipolar nanoparticles (Fig. \ref{fig:2}g-l). In addition to different anisotropy, each molecule has hyperpolarizabilities with different tensor components as given in \cite{Schweig1967a, Schweig1967b}. Our numerical results show that the interaction has similar features as in the simplified model used earlier. The optical nonlinearity strengthens the vdW attraction in most cases, although repulsion is possible for two azulenes providing they are on the $x-$ axis with an opposite direction of their optical axis (Fig. \ref{fig:2}g). 

\begin{figure}[htbp]
   \centering
   \includegraphics[width = \columnwidth]{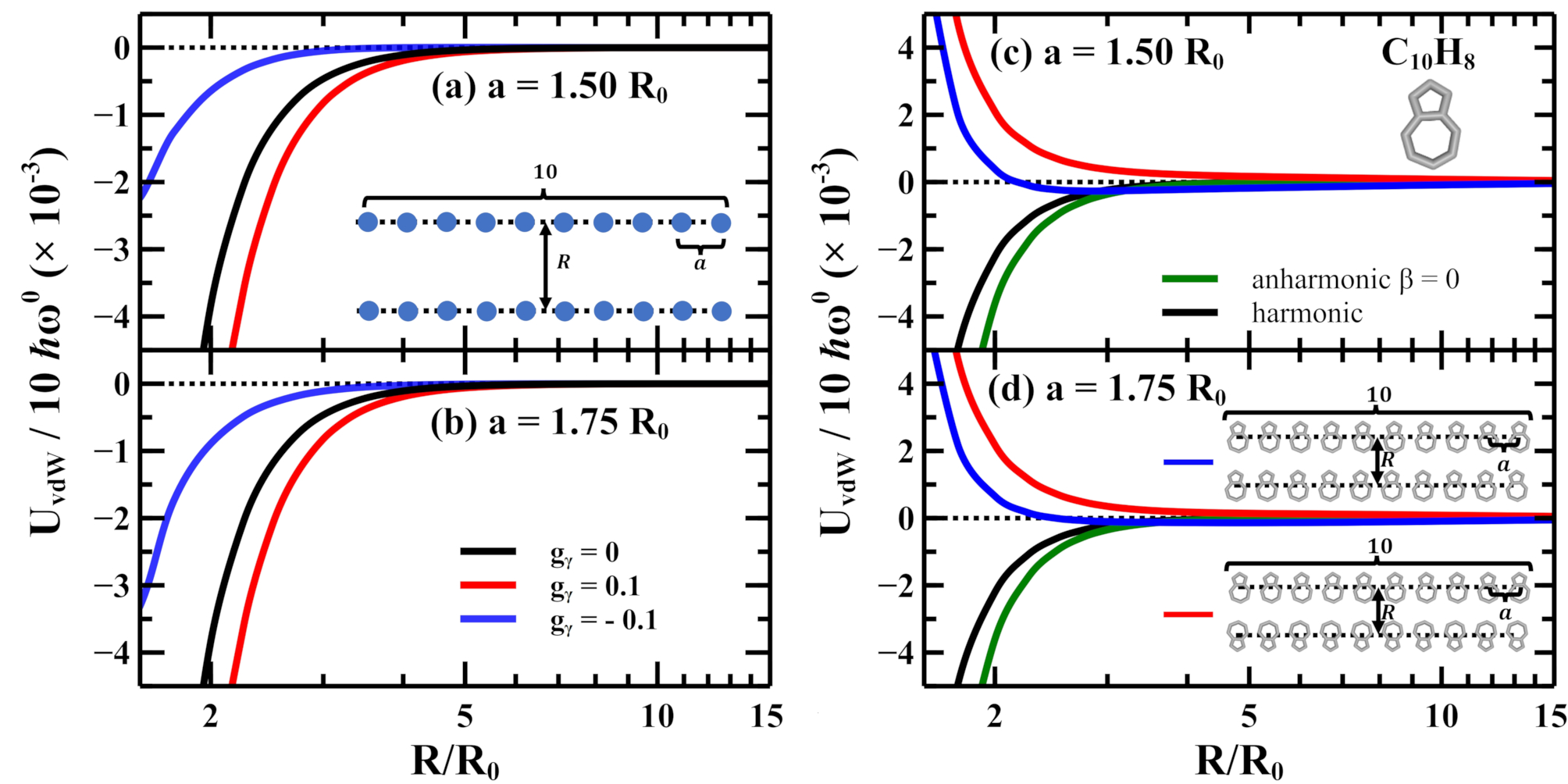}
   \caption{\label{fig:3}  VdW energy rescaled by $\hbar \omega^0$ vs nanoparticle separation rescaled by $R_0=(\alpha/2\pi\epsilon_0)^{1/3}$ is shown in all panels. Two chains where each one has 10 isotropic molecules along $x$-axis and separated by (a) $a=1.5 R_0$; (b) $a=1.75 R_0$ for $g_\gamma=0$ (black), $g_\gamma=0.1$ (red), and $g_\gamma=-0.1$ (blue). Two chains where each one has 10 azulenes along $x$-axis and separated by (a) $a=1.5 R_0$; (b) $a=1.75 R_0$ with $\psi=0$ (red) and $\psi=0$ (blue). Results for $g_\beta=g_\gamma=0$ (black) and $g_\beta=0, g_\gamma\neq 0$ (green) are shown.}
\end{figure}

In addition to the importance of the different optical properties, the vdW interaction is also a many-body phenomenon. The collective nature of dispersive interaction has been examined by several computational methods at the linear response level \cite{Ambrosetti2016, Maurer2019}, however within this approach one can also account for the effect of optical nonlinearities. Using the scheme from Fig.  \ref{fig:flowchart}, we calculate the vdW energy between two chains each with 10 nanoparticles separated by a distance $a$ along the $x$-axis. In the case of isotropic nanoparticles, Fig.  \ref{fig:3}a,b shows that $U_{vdW}$ is attractive but $g_\gamma<0$ reduces its strength, while $g_\gamma>0$ enhances it. In the case of azulene chains whose relative angles are $\psi=0$ and $\pi$, the results given in Fig. \ref{fig:3}d,e show that the interaction is repulsive. This effect is dominated by the second order nonlinearity, since neglecting $g_\beta$ in the simulations turns $U_{vdW}$ into attraction. Note that this is unlike the two-particle interaction (Fig. \ref{fig:2}g) for which the $U_{vdW}$ was found to always be attractive.

\paragraph{Conclusion ---} 
The  \textit{Discrete Coupled Nonlinear Dipole} method presented here establishes a computational framework of vdW interactions that takes into account linear and nonlinear optical properties from a microscopic perspective. This approach relies on a newly derived quantum mechanical Hamiltonian of a nonlinear dipole combined with a perturbation theory treating linear properties as dominant when compared to second and third order optical nonlinearities. Given that vdW interactions are truly collective phenomena, many factors need to be considered for calculating interaction energies in a given system, as shown by advanced computational methods based entirely on linear response theory. Our method gives the means to add another layer of complexity concerning the role of hyperpolarizability in a system with many interacting components. 

Indeed, the second order nonlinear hyperpolarizbility plays an especially prominent in the dispersive coupling. In many cases, it can lead to a significant reduction of the vdW attraction, while in other situations it can result in repulsion. This is a consequence of the complex and un-intuitive interplay between anisotropy, relative position of the nanoparticles, and the particular tensor structure of the optical nonlinearity. Perhaps, the most striking signature of $\mathbf{\beta}$ is its much long-ranged outreach ($1/R^3$) as compared to the characteristic two-body London interaction. It is even possible to tailor the materials properties to completely suppress the $1/R^6$ behavior and have a situation when the energy is completely dominated by the second order nonlinearity. The many-body effects bring additional complexity to the interaction, as found for the azulenes, for example: an attraction between two molecules with $\psi = 0$ can turn into a repulsion between chains of the same molecules.

The newly found quantum mechanical Hamiltonian for nonlinear dipoles and the novel computational schemes for interactions between nonlinear dipoles indeed open up new directions for vdW phenomena and their close relations to materials. In addition to the specific molecules examined here with a diverse structure of their $\bf{\beta}$ and $\bf{\gamma}$ tensors, other particles, such as alkali-halide clusters, have giant polarizabilities resulting in $R_0$ being 3-4 times bigger than the physical size of the cluster \cite{Rayane2002}. For such materials, taking into account the optical nonlinearities may be crucial for the overall stability of the interacting system. This is also relevant to noncentrosymmetric materials with giant optical nonlinearities.  In light of its inter-dependence on $\alpha$, $\beta$, $\gamma$, the vdW interaction and its effects on stacking patterns, interlayer distances, bond lengths, etc... must be re-examined for a more accurate description. Our \textit{Discrete Coupled Nonlinear Dipole} method can easily be applied to such situations and give further qualitative and quantitative understanding of the collective competition of linear and nonlinear many-body effects in vdW interactions in direct relation to equilibrium separations and binding and adhesion energies in materials. This approach can serve as a stepping stone for future design of main-stream computational schemes for more accurate simulations, which might be necessary for nonlinear materials with chemically inert components.

\begin{acknowledgments}
\paragraph{Acknowledgments ---} We acknowledge support from the US Department of Energy under Grant No. DE-FG02-06ER46297. P. R.-L. acknowledges support from AYUDA PUENTE 2022, URJC and the hospitality of the Theory of Light-Matter and Quantum Phenomena group at the Laboratoire Charles Coulomb, University of Montpellier, where part of this work was done. 
\end{acknowledgments}
\bibliography{ref}

\begin{thebibliography}{49}%
\makeatletter
\providecommand \@ifxundefined [1]{%
 \@ifx{#1\undefined}
}%
\providecommand \@ifnum [1]{%
 \ifnum #1\expandafter \@firstoftwo
 \else \expandafter \@secondoftwo
 \fi
}%
\providecommand \@ifx [1]{%
 \ifx #1\expandafter \@firstoftwo
 \else \expandafter \@secondoftwo
 \fi
}%
\providecommand \natexlab [1]{#1}%
\providecommand \enquote  [1]{``#1''}%
\providecommand \bibnamefont  [1]{#1}%
\providecommand \bibfnamefont [1]{#1}%
\providecommand \citenamefont [1]{#1}%
\providecommand \href@noop [0]{\@secondoftwo}%
\providecommand \href [0]{\begingroup \@sanitize@url \@href}%
\providecommand \@href[1]{\@@startlink{#1}\@@href}%
\providecommand \@@href[1]{\endgroup#1\@@endlink}%
\providecommand \@sanitize@url [0]{\catcode `\\12\catcode `\$12\catcode
  `\&12\catcode `\#12\catcode `\^12\catcode `\_12\catcode `\%12\relax}%
\providecommand \@@startlink[1]{}%
\providecommand \@@endlink[0]{}%
\providecommand \url  [0]{\begingroup\@sanitize@url \@url }%
\providecommand \@url [1]{\endgroup\@href {#1}{\urlprefix }}%
\providecommand \urlprefix  [0]{URL }%
\providecommand \Eprint [0]{\href }%
\providecommand \doibase [0]{https://doi.org/}%
\providecommand \selectlanguage [0]{\@gobble}%
\providecommand \bibinfo  [0]{\@secondoftwo}%
\providecommand \bibfield  [0]{\@secondoftwo}%
\providecommand \translation [1]{[#1]}%
\providecommand \BibitemOpen [0]{}%
\providecommand \bibitemStop [0]{}%
\providecommand \bibitemNoStop [0]{.\EOS\space}%
\providecommand \EOS [0]{\spacefactor3000\relax}%
\providecommand \BibitemShut  [1]{\csname bibitem#1\endcsname}%
\let\auto@bib@innerbib\@empty
\bibitem [{\citenamefont {Gjerding}\ \emph {et~al.}(2021)\citenamefont
  {Gjerding}, \citenamefont {Taghizadeh}, \citenamefont {Rasmussen},
  \citenamefont {Ali}, \citenamefont {Bertoldo}, \citenamefont {Deilmann},
  \citenamefont {Knøsgaard}, \citenamefont {Kruse}, \citenamefont {Larsen},
  \citenamefont {Manti}, \citenamefont {Pedersen}, \citenamefont {Petralanda},
  \citenamefont {Skovhus}, \citenamefont {Svendsen}, \citenamefont {Mortensen},
  \citenamefont {Olsen},\ and\ \citenamefont {Thygesen}}]{Gjerding_2021}%
  \BibitemOpen
  \bibfield  {author} {\bibinfo {author} {\bibfnamefont {M.~N.}\ \bibnamefont
  {Gjerding}}, \bibinfo {author} {\bibfnamefont {A.}~\bibnamefont
  {Taghizadeh}}, \bibinfo {author} {\bibfnamefont {A.}~\bibnamefont
  {Rasmussen}}, \bibinfo {author} {\bibfnamefont {S.}~\bibnamefont {Ali}},
  \bibinfo {author} {\bibfnamefont {F.}~\bibnamefont {Bertoldo}}, \bibinfo
  {author} {\bibfnamefont {T.}~\bibnamefont {Deilmann}}, \bibinfo {author}
  {\bibfnamefont {N.~R.}\ \bibnamefont {Knøsgaard}}, \bibinfo {author}
  {\bibfnamefont {M.}~\bibnamefont {Kruse}}, \bibinfo {author} {\bibfnamefont
  {A.~H.}\ \bibnamefont {Larsen}}, \bibinfo {author} {\bibfnamefont
  {S.}~\bibnamefont {Manti}}, \bibinfo {author} {\bibfnamefont {T.~G.}\
  \bibnamefont {Pedersen}}, \bibinfo {author} {\bibfnamefont {U.}~\bibnamefont
  {Petralanda}}, \bibinfo {author} {\bibfnamefont {T.}~\bibnamefont {Skovhus}},
  \bibinfo {author} {\bibfnamefont {M.~K.}\ \bibnamefont {Svendsen}}, \bibinfo
  {author} {\bibfnamefont {J.~J.}\ \bibnamefont {Mortensen}}, \bibinfo {author}
  {\bibfnamefont {T.}~\bibnamefont {Olsen}},\ and\ \bibinfo {author}
  {\bibfnamefont {K.~S.}\ \bibnamefont {Thygesen}},\ }\bibfield  {title}
  {\bibinfo {title} {{Recent progress of the Computational 2D Materials
  Database (C2DB)}},\ }\href {https://doi.org/10.1088/2053-1583/ac1059}
  {\bibfield  {journal} {\bibinfo  {journal} {2D Mater.}\ }\textbf {\bibinfo
  {volume} {8}},\ \bibinfo {pages} {044002} (\bibinfo {year}
  {2021})}\BibitemShut {NoStop}%
\bibitem [{\citenamefont {Geng}\ and\ \citenamefont {Yang}(2018)}]{Geng2018}%
  \BibitemOpen
  \bibfield  {author} {\bibinfo {author} {\bibfnamefont {D.}~\bibnamefont
  {Geng}}\ and\ \bibinfo {author} {\bibfnamefont {H.~Y.}\ \bibnamefont
  {Yang}},\ }\bibfield  {title} {\bibinfo {title} {{Recent Advances in Growth
  of Novel 2D Materials: Beyond Graphene and Transition Metal
  Dichalcogenides}},\ }\href
  {https://doi.org/https://doi.org/10.1002/adma.201800865} {\bibfield
  {journal} {\bibinfo  {journal} {Adv. Mater.}\ }\textbf {\bibinfo {volume}
  {30}},\ \bibinfo {pages} {1800865} (\bibinfo {year} {2018})}\BibitemShut
  {NoStop}%
\bibitem [{\citenamefont {Pandey}\ \emph {et~al.}(2020)\citenamefont {Pandey},
  \citenamefont {Das},\ and\ \citenamefont {Mahadevan}}]{Pandey2020}%
  \BibitemOpen
  \bibfield  {author} {\bibinfo {author} {\bibfnamefont {S.~K.}\ \bibnamefont
  {Pandey}}, \bibinfo {author} {\bibfnamefont {R.}~\bibnamefont {Das}},\ and\
  \bibinfo {author} {\bibfnamefont {P.}~\bibnamefont {Mahadevan}},\ }\bibfield
  {title} {\bibinfo {title} {{Layer-Dependent Electronic Structure Changes in
  Transition Metal Dichalcogenides: The Microscopic Origin}},\ }\href
  {https://doi.org/10.1021/acsomega.0c01138} {\bibfield  {journal} {\bibinfo
  {journal} {ACS Omega}\ }\textbf {\bibinfo {volume} {5}},\ \bibinfo {pages}
  {15169} (\bibinfo {year} {2020})}\BibitemShut {NoStop}%
\bibitem [{\citenamefont {Barik}\ and\ \citenamefont
  {Woods}(2023)}]{Barik2022}%
  \BibitemOpen
  \bibfield  {author} {\bibinfo {author} {\bibfnamefont {R.~K.}\ \bibnamefont
  {Barik}}\ and\ \bibinfo {author} {\bibfnamefont {L.~M.}\ \bibnamefont
  {Woods}},\ }\bibfield  {title} {\bibinfo {title} {{High throughput
  calculations for a dataset of bilayer materials}},\ }\href
  {https://doi.org/10.1038%2Fs41597-023-02146-7} {\bibfield  {journal}
  {\bibinfo  {journal} {Sci. Data}\ }\textbf {\bibinfo {volume} {10}} (\bibinfo
  {year} {2023})}\BibitemShut {NoStop}%
\bibitem [{\citenamefont {Yu}\ \emph {et~al.}(2018)\citenamefont {Yu},
  \citenamefont {Yakobson},\ and\ \citenamefont {Zhang}}]{Yu2018}%
  \BibitemOpen
  \bibfield  {author} {\bibinfo {author} {\bibfnamefont {Z.~G.}\ \bibnamefont
  {Yu}}, \bibinfo {author} {\bibfnamefont {B.~I.}\ \bibnamefont {Yakobson}},\
  and\ \bibinfo {author} {\bibfnamefont {Y.-W.}\ \bibnamefont {Zhang}},\
  }\bibfield  {title} {\bibinfo {title} {{Realizing Indirect-to-Direct Band Gap
  Transition in Few-Layer Two-Dimensional MX2 (M = Mo, W; X = S, Se)}},\ }\href
  {https://doi.org/10.1021/acsaem.8b00774} {\bibfield  {journal} {\bibinfo
  {journal} {ACS Appl. Energy Mater.}\ }\textbf {\bibinfo {volume} {1}},\
  \bibinfo {pages} {4115} (\bibinfo {year} {2018})}\BibitemShut {NoStop}%
\bibitem [{\citenamefont {Ambrosetti}\ \emph {et~al.}(2016)\citenamefont
  {Ambrosetti}, \citenamefont {Ferri}, \citenamefont {DiStasio},\ and\
  \citenamefont {Tkatchenko}}]{Ambrosetti2016}%
  \BibitemOpen
  \bibfield  {author} {\bibinfo {author} {\bibfnamefont {A.}~\bibnamefont
  {Ambrosetti}}, \bibinfo {author} {\bibfnamefont {N.}~\bibnamefont {Ferri}},
  \bibinfo {author} {\bibfnamefont {R.~A.}\ \bibnamefont {DiStasio}},\ and\
  \bibinfo {author} {\bibfnamefont {A.}~\bibnamefont {Tkatchenko}},\ }\bibfield
   {title} {\bibinfo {title} {{Wavelike charge density fluctuations and van der
  Waals interactions at the nanoscale}},\ }\href
  {https://doi.org/10.1126/science.aae0509} {\bibfield  {journal} {\bibinfo
  {journal} {Science}\ }\textbf {\bibinfo {volume} {351}},\ \bibinfo {pages}
  {1171} (\bibinfo {year} {2016})}\BibitemShut {NoStop}%
\bibitem [{\citenamefont {Maurer}\ \emph {et~al.}(2019)\citenamefont {Maurer},
  \citenamefont {Freysoldt}, \citenamefont {Reilly}, \citenamefont
  {Brandenburg}, \citenamefont {Hofmann}, \citenamefont {Bj\"{o}rkman},
  \citenamefont {Leb\`{e}gue},\ and\ \citenamefont {Tkatchenko}}]{Maurer2019}%
  \BibitemOpen
  \bibfield  {author} {\bibinfo {author} {\bibfnamefont {R.~J.}\ \bibnamefont
  {Maurer}}, \bibinfo {author} {\bibfnamefont {C.}~\bibnamefont {Freysoldt}},
  \bibinfo {author} {\bibfnamefont {A.~M.}\ \bibnamefont {Reilly}}, \bibinfo
  {author} {\bibfnamefont {J.~G.}\ \bibnamefont {Brandenburg}}, \bibinfo
  {author} {\bibfnamefont {O.~T.}\ \bibnamefont {Hofmann}}, \bibinfo {author}
  {\bibfnamefont {T.}~\bibnamefont {Bj\"{o}rkman}}, \bibinfo {author}
  {\bibfnamefont {S.}~\bibnamefont {Leb\`{e}gue}},\ and\ \bibinfo {author}
  {\bibfnamefont {A.}~\bibnamefont {Tkatchenko}},\ }\bibfield  {title}
  {\bibinfo {title} {{Advances in Density-Functional Calculations for Materials
  Modeling}},\ }\href {https://doi.org/10.1146/annurev-matsci-070218-010143}
  {\bibfield  {journal} {\bibinfo  {journal} {Ann. Rev. Mater. Res.}\ }\textbf
  {\bibinfo {volume} {49}},\ \bibinfo {pages} {1} (\bibinfo {year}
  {2019})}\BibitemShut {NoStop}%
\bibitem [{\citenamefont {Kresse}\ and\ \citenamefont
  {Furthm\"uller}(1996{\natexlab{a}})}]{Kresse1996}%
  \BibitemOpen
  \bibfield  {author} {\bibinfo {author} {\bibfnamefont {G.}~\bibnamefont
  {Kresse}}\ and\ \bibinfo {author} {\bibfnamefont {J.}~\bibnamefont
  {Furthm\"uller}},\ }\bibfield  {title} {\bibinfo {title} {{Efficiency of
  ab-initio total energy calculations for metals and semiconductors using a
  plane-wave basis set}},\ }\href
  {https://doi.org/https://doi.org/10.1016/0927-0256(96)00008-0} {\bibfield
  {journal} {\bibinfo  {journal} {Comp. Mater. Sci.}\ }\textbf {\bibinfo
  {volume} {6}},\ \bibinfo {pages} {15} (\bibinfo {year}
  {1996}{\natexlab{a}})}\BibitemShut {NoStop}%
\bibitem [{\citenamefont {Kresse}\ and\ \citenamefont
  {Furthm\"uller}(1996{\natexlab{b}})}]{Kresse1996b}%
  \BibitemOpen
  \bibfield  {author} {\bibinfo {author} {\bibfnamefont {G.}~\bibnamefont
  {Kresse}}\ and\ \bibinfo {author} {\bibfnamefont {J.}~\bibnamefont
  {Furthm\"uller}},\ }\bibfield  {title} {\bibinfo {title} {{Efficient
  iterative schemes for ab initio total-energy calculations using a plane-wave
  basis set}},\ }\href {https://doi.org/10.1103/PhysRevB.54.11169} {\bibfield
  {journal} {\bibinfo  {journal} {Phys. Rev. B}\ }\textbf {\bibinfo {volume}
  {54}},\ \bibinfo {pages} {11169} (\bibinfo {year}
  {1996}{\natexlab{b}})}\BibitemShut {NoStop}%
\bibitem [{\citenamefont {Giannozzi}\ \emph {et~al.}(2017)\citenamefont
  {Giannozzi}, \citenamefont {Andreussi}, \citenamefont {Brumme}, \citenamefont
  {Bunau}, \citenamefont {Nardelli}, \citenamefont {Calandra}, \citenamefont
  {Car}, \citenamefont {Cavazzoni}, \citenamefont {Ceresoli}, \citenamefont
  {Cococcioni}, \citenamefont {Colonna}, \citenamefont {Carnimeo},
  \citenamefont {Corso}, \citenamefont {de~Gironcoli}, \citenamefont {Delugas},
  \citenamefont {DiStasio}, \citenamefont {Ferretti}, \citenamefont {Floris},
  \citenamefont {Fratesi}, \citenamefont {Fugallo}, \citenamefont {Gebauer},
  \citenamefont {Gerstmann}, \citenamefont {Giustino}, \citenamefont {Gorni},
  \citenamefont {Jia}, \citenamefont {Kawamura}, \citenamefont {Ko},
  \citenamefont {Kokalj}, \citenamefont {Küçükbenli}, \citenamefont
  {Lazzeri}, \citenamefont {Marsili}, \citenamefont {Marzari}, \citenamefont
  {Mauri}, \citenamefont {Nguyen}, \citenamefont {Nguyen}, \citenamefont {de-la
  Roza}, \citenamefont {Paulatto}, \citenamefont {Poncé}, \citenamefont
  {Rocca}, \citenamefont {Sabatini}, \citenamefont {Santra}, \citenamefont
  {Schlipf}, \citenamefont {Seitsonen}, \citenamefont {Smogunov}, \citenamefont
  {Timrov}, \citenamefont {Thonhauser}, \citenamefont {Umari}, \citenamefont
  {Vast}, \citenamefont {Wu},\ and\ \citenamefont {Baroni}}]{Giannozzi2017}%
  \BibitemOpen
  \bibfield  {author} {\bibinfo {author} {\bibfnamefont {P.}~\bibnamefont
  {Giannozzi}}, \bibinfo {author} {\bibfnamefont {O.}~\bibnamefont
  {Andreussi}}, \bibinfo {author} {\bibfnamefont {T.}~\bibnamefont {Brumme}},
  \bibinfo {author} {\bibfnamefont {O.}~\bibnamefont {Bunau}}, \bibinfo
  {author} {\bibfnamefont {M.~B.}\ \bibnamefont {Nardelli}}, \bibinfo {author}
  {\bibfnamefont {M.}~\bibnamefont {Calandra}}, \bibinfo {author}
  {\bibfnamefont {R.}~\bibnamefont {Car}}, \bibinfo {author} {\bibfnamefont
  {C.}~\bibnamefont {Cavazzoni}}, \bibinfo {author} {\bibfnamefont
  {D.}~\bibnamefont {Ceresoli}}, \bibinfo {author} {\bibfnamefont
  {M.}~\bibnamefont {Cococcioni}}, \bibinfo {author} {\bibfnamefont
  {N.}~\bibnamefont {Colonna}}, \bibinfo {author} {\bibfnamefont
  {I.}~\bibnamefont {Carnimeo}}, \bibinfo {author} {\bibfnamefont {A.~D.}\
  \bibnamefont {Corso}}, \bibinfo {author} {\bibfnamefont {S.}~\bibnamefont
  {de~Gironcoli}}, \bibinfo {author} {\bibfnamefont {P.}~\bibnamefont
  {Delugas}}, \bibinfo {author} {\bibfnamefont {R.~A.}\ \bibnamefont
  {DiStasio}}, \bibinfo {author} {\bibfnamefont {A.}~\bibnamefont {Ferretti}},
  \bibinfo {author} {\bibfnamefont {A.}~\bibnamefont {Floris}}, \bibinfo
  {author} {\bibfnamefont {G.}~\bibnamefont {Fratesi}}, \bibinfo {author}
  {\bibfnamefont {G.}~\bibnamefont {Fugallo}}, \bibinfo {author} {\bibfnamefont
  {R.}~\bibnamefont {Gebauer}}, \bibinfo {author} {\bibfnamefont
  {U.}~\bibnamefont {Gerstmann}}, \bibinfo {author} {\bibfnamefont
  {F.}~\bibnamefont {Giustino}}, \bibinfo {author} {\bibfnamefont
  {T.}~\bibnamefont {Gorni}}, \bibinfo {author} {\bibfnamefont
  {J.}~\bibnamefont {Jia}}, \bibinfo {author} {\bibfnamefont {M.}~\bibnamefont
  {Kawamura}}, \bibinfo {author} {\bibfnamefont {H.-Y.}\ \bibnamefont {Ko}},
  \bibinfo {author} {\bibfnamefont {A.}~\bibnamefont {Kokalj}}, \bibinfo
  {author} {\bibfnamefont {E.}~\bibnamefont {Küçükbenli}}, \bibinfo {author}
  {\bibfnamefont {M.}~\bibnamefont {Lazzeri}}, \bibinfo {author} {\bibfnamefont
  {M.}~\bibnamefont {Marsili}}, \bibinfo {author} {\bibfnamefont
  {N.}~\bibnamefont {Marzari}}, \bibinfo {author} {\bibfnamefont
  {F.}~\bibnamefont {Mauri}}, \bibinfo {author} {\bibfnamefont {N.~L.}\
  \bibnamefont {Nguyen}}, \bibinfo {author} {\bibfnamefont {H.-V.}\
  \bibnamefont {Nguyen}}, \bibinfo {author} {\bibfnamefont {A.~O.}\
  \bibnamefont {de-la Roza}}, \bibinfo {author} {\bibfnamefont
  {L.}~\bibnamefont {Paulatto}}, \bibinfo {author} {\bibfnamefont
  {S.}~\bibnamefont {Poncé}}, \bibinfo {author} {\bibfnamefont
  {D.}~\bibnamefont {Rocca}}, \bibinfo {author} {\bibfnamefont
  {R.}~\bibnamefont {Sabatini}}, \bibinfo {author} {\bibfnamefont
  {B.}~\bibnamefont {Santra}}, \bibinfo {author} {\bibfnamefont
  {M.}~\bibnamefont {Schlipf}}, \bibinfo {author} {\bibfnamefont {A.~P.}\
  \bibnamefont {Seitsonen}}, \bibinfo {author} {\bibfnamefont {A.}~\bibnamefont
  {Smogunov}}, \bibinfo {author} {\bibfnamefont {I.}~\bibnamefont {Timrov}},
  \bibinfo {author} {\bibfnamefont {T.}~\bibnamefont {Thonhauser}}, \bibinfo
  {author} {\bibfnamefont {P.}~\bibnamefont {Umari}}, \bibinfo {author}
  {\bibfnamefont {N.}~\bibnamefont {Vast}}, \bibinfo {author} {\bibfnamefont
  {X.}~\bibnamefont {Wu}},\ and\ \bibinfo {author} {\bibfnamefont
  {S.}~\bibnamefont {Baroni}},\ }\bibfield  {title} {\bibinfo {title}
  {{Advanced capabilities for materials modelling with Quantum ESPRESSO}},\
  }\href {https://doi.org/10.1088/1361-648X/aa8f79} {\bibfield  {journal}
  {\bibinfo  {journal} {J. Phys. Condens. Mat.}\ }\textbf {\bibinfo {volume}
  {29}},\ \bibinfo {pages} {465901} (\bibinfo {year} {2017})}\BibitemShut
  {NoStop}%
\bibitem [{\citenamefont {Li}\ \emph {et~al.}(2013)\citenamefont {Li},
  \citenamefont {Rao}, \citenamefont {Mak}, \citenamefont {You}, \citenamefont
  {Wang}, \citenamefont {Dean},\ and\ \citenamefont {Heinz}}]{Li2013}%
  \BibitemOpen
  \bibfield  {author} {\bibinfo {author} {\bibfnamefont {Y.}~\bibnamefont
  {Li}}, \bibinfo {author} {\bibfnamefont {Y.}~\bibnamefont {Rao}}, \bibinfo
  {author} {\bibfnamefont {K.~F.}\ \bibnamefont {Mak}}, \bibinfo {author}
  {\bibfnamefont {Y.}~\bibnamefont {You}}, \bibinfo {author} {\bibfnamefont
  {S.}~\bibnamefont {Wang}}, \bibinfo {author} {\bibfnamefont {C.~R.}\
  \bibnamefont {Dean}},\ and\ \bibinfo {author} {\bibfnamefont {T.~F.}\
  \bibnamefont {Heinz}},\ }\bibfield  {title} {\bibinfo {title} {{Probing
  Symmetry Properties of Few-Layer MoS2 and h-BN by Optical Second-Harmonic
  Generation}},\ }\href {https://doi.org/10.1021/nl401561r} {\bibfield
  {journal} {\bibinfo  {journal} {Nano Lett.}\ }\textbf {\bibinfo {volume}
  {13}},\ \bibinfo {pages} {3329} (\bibinfo {year} {2013})}\BibitemShut
  {NoStop}%
\bibitem [{\citenamefont {Wang}\ and\ \citenamefont {Guo}(2015)}]{Wang2015}%
  \BibitemOpen
  \bibfield  {author} {\bibinfo {author} {\bibfnamefont {C.-Y.}\ \bibnamefont
  {Wang}}\ and\ \bibinfo {author} {\bibfnamefont {G.-Y.}\ \bibnamefont {Guo}},\
  }\bibfield  {title} {\bibinfo {title} {{Nonlinear Optical Properties of
  Transition-Metal Dichalcogenide MX2 (M = Mo, W; X = S, Se) Monolayers and
  Trilayers from First-Principles Calculations}},\ }\href
  {https://doi.org/10.1021/acs.jpcc.5b01866} {\bibfield  {journal} {\bibinfo
  {journal} {J. Phys. Chem. C}\ }\textbf {\bibinfo {volume} {119}},\ \bibinfo
  {pages} {13268} (\bibinfo {year} {2015})}\BibitemShut {NoStop}%
\bibitem [{\citenamefont {Autere}\ \emph {et~al.}(2018)\citenamefont {Autere},
  \citenamefont {Jussila}, \citenamefont {Marini}, \citenamefont {Saavedra},
  \citenamefont {Dai}, \citenamefont {S\"ayn\"atjoki}, \citenamefont
  {Karvonen}, \citenamefont {Yang}, \citenamefont {Amirsolaimani},
  \citenamefont {Norwood}, \citenamefont {Peyghambarian}, \citenamefont
  {Lipsanen}, \citenamefont {Kieu}, \citenamefont {de~Abajo},\ and\
  \citenamefont {Sun}}]{Autere2018}%
  \BibitemOpen
  \bibfield  {author} {\bibinfo {author} {\bibfnamefont {A.}~\bibnamefont
  {Autere}}, \bibinfo {author} {\bibfnamefont {H.}~\bibnamefont {Jussila}},
  \bibinfo {author} {\bibfnamefont {A.}~\bibnamefont {Marini}}, \bibinfo
  {author} {\bibfnamefont {J.~R.~M.}\ \bibnamefont {Saavedra}}, \bibinfo
  {author} {\bibfnamefont {Y.}~\bibnamefont {Dai}}, \bibinfo {author}
  {\bibfnamefont {A.}~\bibnamefont {S\"ayn\"atjoki}}, \bibinfo {author}
  {\bibfnamefont {L.}~\bibnamefont {Karvonen}}, \bibinfo {author}
  {\bibfnamefont {H.}~\bibnamefont {Yang}}, \bibinfo {author} {\bibfnamefont
  {B.}~\bibnamefont {Amirsolaimani}}, \bibinfo {author} {\bibfnamefont {R.~A.}\
  \bibnamefont {Norwood}}, \bibinfo {author} {\bibfnamefont {N.}~\bibnamefont
  {Peyghambarian}}, \bibinfo {author} {\bibfnamefont {H.}~\bibnamefont
  {Lipsanen}}, \bibinfo {author} {\bibfnamefont {K.}~\bibnamefont {Kieu}},
  \bibinfo {author} {\bibfnamefont {F.~J.~G.}\ \bibnamefont {de~Abajo}},\ and\
  \bibinfo {author} {\bibfnamefont {Z.}~\bibnamefont {Sun}},\ }\bibfield
  {title} {\bibinfo {title} {{Optical harmonic generation in monolayer group-VI
  transition metal dichalcogenides}},\ }\href
  {https://doi.org/10.1103/PhysRevB.98.115426} {\bibfield  {journal} {\bibinfo
  {journal} {Phys. Rev. B}\ }\textbf {\bibinfo {volume} {98}},\ \bibinfo
  {pages} {115426} (\bibinfo {year} {2018})}\BibitemShut {NoStop}%
\bibitem [{\citenamefont {Wen}\ \emph {et~al.}(2019)\citenamefont {Wen},
  \citenamefont {Gong},\ and\ \citenamefont {Li}}]{Wen2019}%
  \BibitemOpen
  \bibfield  {author} {\bibinfo {author} {\bibfnamefont {X.}~\bibnamefont
  {Wen}}, \bibinfo {author} {\bibfnamefont {Z.}~\bibnamefont {Gong}},\ and\
  \bibinfo {author} {\bibfnamefont {D.}~\bibnamefont {Li}},\ }\bibfield
  {title} {\bibinfo {title} {{Nonlinear optics of two-dimensional transition
  metal dichalcogenides}},\ }\href
  {https://doi.org/https://doi.org/10.1002/inf2.12024} {\bibfield  {journal}
  {\bibinfo  {journal} {InfoMat}\ }\textbf {\bibinfo {volume} {1}},\ \bibinfo
  {pages} {317} (\bibinfo {year} {2019})}\BibitemShut {NoStop}%
\bibitem [{\citenamefont {Patankar}\ \emph {et~al.}(2018)\citenamefont
  {Patankar}, \citenamefont {Wu}, \citenamefont {Lu}, \citenamefont {Rai},
  \citenamefont {Tran}, \citenamefont {Morimoto}, \citenamefont {Parker},
  \citenamefont {Grushin}, \citenamefont {Nair}, \citenamefont {Analytis},
  \citenamefont {Moore}, \citenamefont {Orenstein},\ and\ \citenamefont
  {Torchinsky}}]{Patankar2018}%
  \BibitemOpen
  \bibfield  {author} {\bibinfo {author} {\bibfnamefont {S.}~\bibnamefont
  {Patankar}}, \bibinfo {author} {\bibfnamefont {L.}~\bibnamefont {Wu}},
  \bibinfo {author} {\bibfnamefont {B.}~\bibnamefont {Lu}}, \bibinfo {author}
  {\bibfnamefont {M.}~\bibnamefont {Rai}}, \bibinfo {author} {\bibfnamefont
  {J.~D.}\ \bibnamefont {Tran}}, \bibinfo {author} {\bibfnamefont
  {T.}~\bibnamefont {Morimoto}}, \bibinfo {author} {\bibfnamefont {D.~E.}\
  \bibnamefont {Parker}}, \bibinfo {author} {\bibfnamefont {A.~G.}\
  \bibnamefont {Grushin}}, \bibinfo {author} {\bibfnamefont {N.~L.}\
  \bibnamefont {Nair}}, \bibinfo {author} {\bibfnamefont {J.~G.}\ \bibnamefont
  {Analytis}}, \bibinfo {author} {\bibfnamefont {J.~E.}\ \bibnamefont {Moore}},
  \bibinfo {author} {\bibfnamefont {J.}~\bibnamefont {Orenstein}},\ and\
  \bibinfo {author} {\bibfnamefont {D.~H.}\ \bibnamefont {Torchinsky}},\
  }\bibfield  {title} {\bibinfo {title} {{Resonance-enhanced optical
  nonlinearity in the Weyl semimetal TaAs}},\ }\href
  {https://doi.org/10.1103/PhysRevB.98.165113} {\bibfield  {journal} {\bibinfo
  {journal} {Phys. Rev. B}\ }\textbf {\bibinfo {volume} {98}},\ \bibinfo
  {pages} {165113} (\bibinfo {year} {2018})}\BibitemShut {NoStop}%
\bibitem [{\citenamefont {Takasan}\ \emph {et~al.}(2021)\citenamefont
  {Takasan}, \citenamefont {Morimoto}, \citenamefont {Orenstein},\ and\
  \citenamefont {Moore}}]{Takasan2021}%
  \BibitemOpen
  \bibfield  {author} {\bibinfo {author} {\bibfnamefont {K.}~\bibnamefont
  {Takasan}}, \bibinfo {author} {\bibfnamefont {T.}~\bibnamefont {Morimoto}},
  \bibinfo {author} {\bibfnamefont {J.}~\bibnamefont {Orenstein}},\ and\
  \bibinfo {author} {\bibfnamefont {J.~E.}\ \bibnamefont {Moore}},\ }\bibfield
  {title} {\bibinfo {title} {{Current-induced second harmonic generation in
  inversion-symmetric Dirac and Weyl semimetals}},\ }\href
  {https://doi.org/10.1103/PhysRevB.104.L161202} {\bibfield  {journal}
  {\bibinfo  {journal} {Phys. Rev. B}\ }\textbf {\bibinfo {volume} {104}},\
  \bibinfo {pages} {L161202} (\bibinfo {year} {2021})}\BibitemShut {NoStop}%
\bibitem [{\citenamefont {Drueke}\ \emph {et~al.}(2021)\citenamefont {Drueke},
  \citenamefont {Yang},\ and\ \citenamefont {Zhao}}]{Drueke2021}%
  \BibitemOpen
  \bibfield  {author} {\bibinfo {author} {\bibfnamefont {E.}~\bibnamefont
  {Drueke}}, \bibinfo {author} {\bibfnamefont {J.}~\bibnamefont {Yang}},\ and\
  \bibinfo {author} {\bibfnamefont {L.}~\bibnamefont {Zhao}},\ }\bibfield
  {title} {\bibinfo {title} {{Observation of strong and anisotropic nonlinear
  optical effects through polarization-resolved optical spectroscopy in the
  type-II Weyl semimetal
  ${T}_{\mathrm{d}}\text{\ensuremath{-}}\mathrm{W}{\mathrm{Te}}_{2}$}},\ }\href
  {https://doi.org/10.1103/PhysRevB.104.064304} {\bibfield  {journal} {\bibinfo
   {journal} {Phys. Rev. B}\ }\textbf {\bibinfo {volume} {104}},\ \bibinfo
  {pages} {064304} (\bibinfo {year} {2021})}\BibitemShut {NoStop}%
\bibitem [{\citenamefont {Xu}\ \emph {et~al.}(2020)\citenamefont {Xu},
  \citenamefont {Zhang}, \citenamefont {Koepernik}, \citenamefont {Shi},
  \citenamefont {van~den Brink}, \citenamefont {Felser},\ and\ \citenamefont
  {Sun}}]{Xu2020}%
  \BibitemOpen
  \bibfield  {author} {\bibinfo {author} {\bibfnamefont {Q.}~\bibnamefont
  {Xu}}, \bibinfo {author} {\bibfnamefont {Y.}~\bibnamefont {Zhang}}, \bibinfo
  {author} {\bibfnamefont {K.}~\bibnamefont {Koepernik}}, \bibinfo {author}
  {\bibfnamefont {W.}~\bibnamefont {Shi}}, \bibinfo {author} {\bibfnamefont
  {J.}~\bibnamefont {van~den Brink}}, \bibinfo {author} {\bibfnamefont
  {C.}~\bibnamefont {Felser}},\ and\ \bibinfo {author} {\bibfnamefont
  {Y.}~\bibnamefont {Sun}},\ }\bibfield  {title} {\bibinfo {title}
  {{Comprehensive scan for nonmagnetic Weyl semimetals with nonlinear optical
  response}},\ }\href {https://doi.org/10.1038/s41524-020-0301-1} {\bibfield
  {journal} {\bibinfo  {journal} {npj Comput. Mater.}\ }\textbf {\bibinfo
  {volume} {6}},\ \bibinfo {pages} {32} (\bibinfo {year} {2020})}\BibitemShut
  {NoStop}%
\bibitem [{\citenamefont {You}\ \emph {et~al.}(2019)\citenamefont {You},
  \citenamefont {Bongu}, \citenamefont {Bao},\ and\ \citenamefont
  {Panoiu}}]{You2019}%
  \BibitemOpen
  \bibfield  {author} {\bibinfo {author} {\bibfnamefont {J.}~\bibnamefont
  {You}}, \bibinfo {author} {\bibfnamefont {S.}~\bibnamefont {Bongu}}, \bibinfo
  {author} {\bibfnamefont {Q.}~\bibnamefont {Bao}},\ and\ \bibinfo {author}
  {\bibfnamefont {N.}~\bibnamefont {Panoiu}},\ }\bibfield  {title} {\bibinfo
  {title} {{Nonlinear optical properties and applications of 2D materials:
  theoretical and experimental aspects}},\ }\href
  {https://doi.org/doi:10.1515/nanoph-2018-0106} {\bibfield  {journal}
  {\bibinfo  {journal} {Nanophotonics}\ }\textbf {\bibinfo {volume} {8}},\
  \bibinfo {pages} {63} (\bibinfo {year} {2019})}\BibitemShut {NoStop}%
\bibitem [{\citenamefont {Zhou}\ \emph {et~al.}(2020)\citenamefont {Zhou},
  \citenamefont {Fu}, \citenamefont {Lv}, \citenamefont {Wang}, \citenamefont
  {Gao}, \citenamefont {Li}, \citenamefont {Deng},\ and\ \citenamefont
  {Xiong}}]{Zhou2020}%
  \BibitemOpen
  \bibfield  {author} {\bibinfo {author} {\bibfnamefont {L.}~\bibnamefont
  {Zhou}}, \bibinfo {author} {\bibfnamefont {H.}~\bibnamefont {Fu}}, \bibinfo
  {author} {\bibfnamefont {T.}~\bibnamefont {Lv}}, \bibinfo {author}
  {\bibfnamefont {C.}~\bibnamefont {Wang}}, \bibinfo {author} {\bibfnamefont
  {H.}~\bibnamefont {Gao}}, \bibinfo {author} {\bibfnamefont {D.}~\bibnamefont
  {Li}}, \bibinfo {author} {\bibfnamefont {L.}~\bibnamefont {Deng}},\ and\
  \bibinfo {author} {\bibfnamefont {W.}~\bibnamefont {Xiong}},\ }\bibfield
  {title} {\bibinfo {title} {{Nonlinear Optical Characterization of 2D
  Materials}},\ }\href {https://www.mdpi.com/2079-4991/10/11/2263} {\bibfield
  {journal} {\bibinfo  {journal} {Nanomater.}\ }\textbf {\bibinfo {volume}
  {10}},\ \bibinfo {pages} {2263} (\bibinfo {year} {2020})}\BibitemShut
  {NoStop}%
\bibitem [{\citenamefont {Soo}\ and\ \citenamefont
  {Kr{\"u}ger}(2016)}]{Soo2016}%
  \BibitemOpen
  \bibfield  {author} {\bibinfo {author} {\bibfnamefont {H.}~\bibnamefont
  {Soo}}\ and\ \bibinfo {author} {\bibfnamefont {M.}~\bibnamefont
  {Kr{\"u}ger}},\ }\bibfield  {title} {\bibinfo {title} {{Fluctuational
  electrodynamics for nonlinear media}},\ }\href
  {https://doi.org/10.1209/0295-5075/115/41002} {\bibfield  {journal} {\bibinfo
   {journal} {Europhys. Lett.}\ }\textbf {\bibinfo {volume} {115}},\ \bibinfo
  {pages} {41002} (\bibinfo {year} {2016})}\BibitemShut {NoStop}%
\bibitem [{\citenamefont {Soo}\ and\ \citenamefont {Kr\"uger}(2018)}]{Soo2018}%
  \BibitemOpen
  \bibfield  {author} {\bibinfo {author} {\bibfnamefont {H.}~\bibnamefont
  {Soo}}\ and\ \bibinfo {author} {\bibfnamefont {M.}~\bibnamefont {Kr\"uger}},\
  }\bibfield  {title} {\bibinfo {title} {{Fluctuational electrodynamics for
  nonlinear materials in and out of thermal equilibrium}},\ }\href
  {https://doi.org/10.1103/PhysRevB.97.045412} {\bibfield  {journal} {\bibinfo
  {journal} {Phys. Rev. B}\ }\textbf {\bibinfo {volume} {97}},\ \bibinfo
  {pages} {045412} (\bibinfo {year} {2018})}\BibitemShut {NoStop}%
\bibitem [{\citenamefont {Soo}\ \emph {et~al.}(2017)\citenamefont {Soo},
  \citenamefont {Dean},\ and\ \citenamefont {Kr\"uger}}]{Soo2017}%
  \BibitemOpen
  \bibfield  {author} {\bibinfo {author} {\bibfnamefont {H.}~\bibnamefont
  {Soo}}, \bibinfo {author} {\bibfnamefont {D.~S.}\ \bibnamefont {Dean}},\ and\
  \bibinfo {author} {\bibfnamefont {M.}~\bibnamefont {Kr\"uger}},\ }\bibfield
  {title} {\bibinfo {title} {{Particles with nonlinear electric response:
  Suppressing van der Waals forces by an external field}},\ }\href
  {https://doi.org/10.1103/PhysRevE.95.012151} {\bibfield  {journal} {\bibinfo
  {journal} {Phys. Rev. E}\ }\textbf {\bibinfo {volume} {95}},\ \bibinfo
  {pages} {012151} (\bibinfo {year} {2017})}\BibitemShut {NoStop}%
\bibitem [{\citenamefont {Massa}\ \emph {et~al.}(2021)\citenamefont {Massa},
  \citenamefont {Ambrosetti},\ and\ \citenamefont {Silvestrelli}}]{Massa2021}%
  \BibitemOpen
  \bibfield  {author} {\bibinfo {author} {\bibfnamefont {D.}~\bibnamefont
  {Massa}}, \bibinfo {author} {\bibfnamefont {A.}~\bibnamefont {Ambrosetti}},\
  and\ \bibinfo {author} {\bibfnamefont {P.~L.}\ \bibnamefont {Silvestrelli}},\
  }\bibfield  {title} {\bibinfo {title} {{Many-body van der Waals interactions
  beyond the dipole approximation}},\ }\href
  {https://doi.org/10.1063/5.0051604} {\bibfield  {journal} {\bibinfo
  {journal} {J. Chem. Phys.}\ }\textbf {\bibinfo {volume} {154}},\ \bibinfo
  {pages} {224115} (\bibinfo {year} {2021})}\BibitemShut {NoStop}%
\bibitem [{\citenamefont {Stedman}\ \emph {et~al.}(2014)\citenamefont
  {Stedman}, \citenamefont {Drosdoff},\ and\ \citenamefont
  {Woods}}]{Stedman2014}%
  \BibitemOpen
  \bibfield  {author} {\bibinfo {author} {\bibfnamefont {T.}~\bibnamefont
  {Stedman}}, \bibinfo {author} {\bibfnamefont {D.}~\bibnamefont {Drosdoff}},\
  and\ \bibinfo {author} {\bibfnamefont {L.~M.}\ \bibnamefont {Woods}},\
  }\bibfield  {title} {\bibinfo {title} {{van der Waals interactions between
  nanostructures: Some analytic results from series expansions}},\ }\href
  {https://doi.org/10.1103/PhysRevA.89.012509} {\bibfield  {journal} {\bibinfo
  {journal} {Phys. Rev. A}\ }\textbf {\bibinfo {volume} {89}},\ \bibinfo
  {pages} {012509} (\bibinfo {year} {2014})}\BibitemShut {NoStop}%
\bibitem [{\citenamefont {Hopkins}\ \emph {et~al.}(2014)\citenamefont
  {Hopkins}, \citenamefont {Dryden}, \citenamefont {Ching}, \citenamefont
  {French}, \citenamefont {Parsegian},\ and\ \citenamefont
  {Podgornik}}]{Hopkins2014}%
  \BibitemOpen
  \bibfield  {author} {\bibinfo {author} {\bibfnamefont {J.~C.}\ \bibnamefont
  {Hopkins}}, \bibinfo {author} {\bibfnamefont {D.~M.}\ \bibnamefont {Dryden}},
  \bibinfo {author} {\bibfnamefont {W.-Y.}\ \bibnamefont {Ching}}, \bibinfo
  {author} {\bibfnamefont {R.~H.}\ \bibnamefont {French}}, \bibinfo {author}
  {\bibfnamefont {V.~A.}\ \bibnamefont {Parsegian}},\ and\ \bibinfo {author}
  {\bibfnamefont {R.}~\bibnamefont {Podgornik}},\ }\bibfield  {title} {\bibinfo
  {title} {{Dielectric response variation and the strength of van der Waals
  interactions}},\ }\href
  {https://doi.org/https://doi.org/10.1016/j.jcis.2013.10.040} {\bibfield
  {journal} {\bibinfo  {journal} {J. Colloid Interface Sci.}\ }\textbf
  {\bibinfo {volume} {417}},\ \bibinfo {pages} {278} (\bibinfo {year}
  {2014})}\BibitemShut {NoStop}%
\bibitem [{\citenamefont {Cole}\ \emph {et~al.}(2009)\citenamefont {Cole},
  \citenamefont {Velegol}, \citenamefont {Kim},\ and\ \citenamefont
  {Lucas}}]{Cole2009}%
  \BibitemOpen
  \bibfield  {author} {\bibinfo {author} {\bibfnamefont {M.~W.}\ \bibnamefont
  {Cole}}, \bibinfo {author} {\bibfnamefont {D.}~\bibnamefont {Velegol}},
  \bibinfo {author} {\bibfnamefont {H.-Y.}\ \bibnamefont {Kim}},\ and\ \bibinfo
  {author} {\bibfnamefont {A.~A.}\ \bibnamefont {Lucas}},\ }\bibfield  {title}
  {\bibinfo {title} {{Nanoscale van der Waals interactions}},\ }\href
  {https://doi.org/10.1080/08927020902929794} {\bibfield  {journal} {\bibinfo
  {journal} {Mol. Simul.}\ }\textbf {\bibinfo {volume} {35}},\ \bibinfo {pages}
  {849} (\bibinfo {year} {2009})}\BibitemShut {NoStop}%
\bibitem [{\citenamefont {Gobre}\ and\ \citenamefont
  {Tkatchenko}(2013)}]{Tkatchenko2013b}%
  \BibitemOpen
  \bibfield  {author} {\bibinfo {author} {\bibfnamefont {V.~V.}\ \bibnamefont
  {Gobre}}\ and\ \bibinfo {author} {\bibfnamefont {A.}~\bibnamefont
  {Tkatchenko}},\ }\bibfield  {title} {\bibinfo {title} {{Scaling laws for van
  der Waals interactions in nanostructured materials}},\ }\href
  {https://doi.org/10.1038/ncomms3341} {\bibfield  {journal} {\bibinfo
  {journal} {Nat. Commun.}\ }\textbf {\bibinfo {volume} {4}},\ \bibinfo {pages}
  {2341} (\bibinfo {year} {2013})}\BibitemShut {NoStop}%
\bibitem [{\citenamefont {Sarabadani}\ \emph {et~al.}(2011)\citenamefont
  {Sarabadani}, \citenamefont {Naji}, \citenamefont {Asgari},\ and\
  \citenamefont {Podgornik}}]{Sarabadani2011}%
  \BibitemOpen
  \bibfield  {author} {\bibinfo {author} {\bibfnamefont {J.}~\bibnamefont
  {Sarabadani}}, \bibinfo {author} {\bibfnamefont {A.}~\bibnamefont {Naji}},
  \bibinfo {author} {\bibfnamefont {R.}~\bibnamefont {Asgari}},\ and\ \bibinfo
  {author} {\bibfnamefont {R.}~\bibnamefont {Podgornik}},\ }\bibfield  {title}
  {\bibinfo {title} {{Many-body effects in the van der Waals--Casimir
  interaction between graphene layers}},\ }\href
  {https://doi.org/10.1103/PhysRevB.84.155407} {\bibfield  {journal} {\bibinfo
  {journal} {Phys. Rev. B}\ }\textbf {\bibinfo {volume} {84}},\ \bibinfo
  {pages} {155407} (\bibinfo {year} {2011})}\BibitemShut {NoStop}%
\bibitem [{\citenamefont {Drosdoff}\ and\ \citenamefont
  {Woods}(2014)}]{Drosdoff2014}%
  \BibitemOpen
  \bibfield  {author} {\bibinfo {author} {\bibfnamefont {D.}~\bibnamefont
  {Drosdoff}}\ and\ \bibinfo {author} {\bibfnamefont {L.~M.}\ \bibnamefont
  {Woods}},\ }\bibfield  {title} {\bibinfo {title} {{Quantum and Thermal
  Dispersion Forces: Application to Graphene Nanoribbons}},\ }\href
  {https://doi.org/10.1103/PhysRevLett.112.025501} {\bibfield  {journal}
  {\bibinfo  {journal} {Phys. Rev. Lett.}\ }\textbf {\bibinfo {volume} {112}},\
  \bibinfo {pages} {025501} (\bibinfo {year} {2014})}\BibitemShut {NoStop}%
\bibitem [{\citenamefont {Lozovskii}\ and\ \citenamefont
  {Khudik}(1990{\natexlab{a}})}]{Lozovskii1990}%
  \BibitemOpen
  \bibfield  {author} {\bibinfo {author} {\bibfnamefont {V.~Z.}\ \bibnamefont
  {Lozovskii}}\ and\ \bibinfo {author} {\bibfnamefont {B.~I.}\ \bibnamefont
  {Khudik}},\ }\bibfield  {title} {\bibinfo {title} {{The New Mechanism of
  Physical Adsorption on Solid Surface. I. Adsorption of Nonpolar Molecules}},\
  }\href {https://doi.org/10.1002/pssb.2221580212} {\bibfield  {journal}
  {\bibinfo  {journal} {Phys. Stat. Sol. (b)}\ }\textbf {\bibinfo {volume}
  {158}},\ \bibinfo {pages} {511} (\bibinfo {year}
  {1990}{\natexlab{a}})}\BibitemShut {NoStop}%
\bibitem [{\citenamefont {Lozovskii}\ and\ \citenamefont
  {Khudik}(1990{\natexlab{b}})}]{Lozovskii1990b}%
  \BibitemOpen
  \bibfield  {author} {\bibinfo {author} {\bibfnamefont {V.~Z.}\ \bibnamefont
  {Lozovskii}}\ and\ \bibinfo {author} {\bibfnamefont {B.~I.}\ \bibnamefont
  {Khudik}},\ }\bibfield  {title} {\bibinfo {title} {{The New Mechanism of
  Physical Adsorption on Solid Surface. II. Adsorption of Polar Molecules}},\
  }\href {https://doi.org/10.1002/pssb.2221600110} {\bibfield  {journal}
  {\bibinfo  {journal} {Phys. Stat. Sol. (b)}\ }\textbf {\bibinfo {volume}
  {160}},\ \bibinfo {pages} {137} (\bibinfo {year}
  {1990}{\natexlab{b}})}\BibitemShut {NoStop}%
\bibitem [{\citenamefont {Makhnovets}\ and\ \citenamefont
  {Kolezhuk}(2016)}]{Makhnovets2016}%
  \BibitemOpen
  \bibfield  {author} {\bibinfo {author} {\bibfnamefont {K.}~\bibnamefont
  {Makhnovets}}\ and\ \bibinfo {author} {\bibfnamefont {A.}~\bibnamefont
  {Kolezhuk}},\ }\bibfield  {title} {\bibinfo {title} {{On short-range
  enhancement of Van-der-Waals forces}},\ }\href
  {https://doi.org/10.1002/mawe.201600461} {\bibfield  {journal} {\bibinfo
  {journal} {Materialwiss. Werkstofftech.}\ }\textbf {\bibinfo {volume} {47}},\
  \bibinfo {pages} {222} (\bibinfo {year} {2016})}\BibitemShut {NoStop}%
\bibitem [{\citenamefont {Boyd}(2007)}]{Boyd2007}%
  \BibitemOpen
  \bibfield  {author} {\bibinfo {author} {\bibfnamefont {R.~W.}\ \bibnamefont
  {Boyd}},\ }\href@noop {} {\emph {\bibinfo {title} {{Nonlinear Optics}}}}\
  (\bibinfo  {publisher} {Academic Press},\ \bibinfo {address} {Rochester,
  NY},\ \bibinfo {year} {2007})\BibitemShut {NoStop}%
\bibitem [{sup()}]{supplementary}%
  \BibitemOpen
  \href@noop {} {}\bibinfo {howpublished} {See Supplementary Information for
  detailed derivations of the Hamiltonian, \textit{Discrete Coupled Nonlinear
  Dipole method} and analytical expressions of two-body vdW interactions; see
  also Ref. \cite{Hoangbook2015}.}\BibitemShut {Stop}%
\bibitem [{\citenamefont {Balla}\ \emph {et~al.}(2010)\citenamefont {Balla},
  \citenamefont {So},\ and\ \citenamefont {Sheppard}}]{Balla2010}%
  \BibitemOpen
  \bibfield  {author} {\bibinfo {author} {\bibfnamefont {N.~K.}\ \bibnamefont
  {Balla}}, \bibinfo {author} {\bibfnamefont {P.~T.~C.}\ \bibnamefont {So}},\
  and\ \bibinfo {author} {\bibfnamefont {C.~J.~R.}\ \bibnamefont {Sheppard}},\
  }\bibfield  {title} {\bibinfo {title} {{Second harmonic scattering from small
  particles using Discrete Dipole Approximation}},\ }\href
  {https://doi.org/10.1364/oe.18.021603} {\bibfield  {journal} {\bibinfo
  {journal} {Opt. Express}\ }\textbf {\bibinfo {volume} {18}},\ \bibinfo
  {pages} {21603} (\bibinfo {year} {2010})}\BibitemShut {NoStop}%
\bibitem [{\citenamefont {Balla}\ \emph {et~al.}(2012)\citenamefont {Balla},
  \citenamefont {Yew}, \citenamefont {Sheppard},\ and\ \citenamefont
  {So}}]{Balla2012}%
  \BibitemOpen
  \bibfield  {author} {\bibinfo {author} {\bibfnamefont {N.~K.}\ \bibnamefont
  {Balla}}, \bibinfo {author} {\bibfnamefont {E.~Y.~S.}\ \bibnamefont {Yew}},
  \bibinfo {author} {\bibfnamefont {C.~J.~R.}\ \bibnamefont {Sheppard}},\ and\
  \bibinfo {author} {\bibfnamefont {P.~T.~C.}\ \bibnamefont {So}},\ }\bibfield
  {title} {\bibinfo {title} {{Coupled and uncoupled dipole models of nonlinear
  scattering}},\ }\href {https://doi.org/10.1364/OE.20.025834} {\bibfield
  {journal} {\bibinfo  {journal} {Opt. Express}\ }\textbf {\bibinfo {volume}
  {20}},\ \bibinfo {pages} {25834} (\bibinfo {year} {2012})}\BibitemShut
  {NoStop}%
\bibitem [{Note1()}]{Note1}%
  \BibitemOpen
  \bibinfo {note} {$E_0 = \protect \frac {1}{2} \protect \mathbb {A}_{mn}
  {P}_{0,m} {P}_{0,n} + \protect \frac {1}{3} \protect \mathbb {B}_{mnq}
  {P}_{0,m} {P}_{0,n} {P}_{0,q} + \protect \frac {1}{4} \protect \mathbb
  {G}_{mnqp} {P}_{0,m} {P}_{0,n} {P}_{0,q} {P}_{0,p}$, $\protect \hat {H}_{h} =
  \protect \frac {1}{2} \left [ \protect \mathbb {A}_{mn} + 2 \protect \mathbb
  {B}_{mnq} P_{0,q} + 3 \protect \mathbb {G}_{mnqp} P_{0,q} P_{0,p} \right ]
  \protect \hat {Q}_m \protect \hat {Q}_n$ and $\protect \hat {H}_{anh}^{\prime
  } = \protect \frac {1}{3} \left ( \protect \mathbb {B}_{mnq} + 3 \protect
  \mathbb {G}_{mnqp} P_{0,p} \right ) \protect \hat {Q}_{m} \protect \hat
  {Q}_{n} \protect \hat {Q}_{q} + \protect \frac {1}{4} \protect \mathbb
  {G}_{mnqp} \protect \hat {Q}_{m} \protect \hat {Q}_{n} \protect \hat {Q}_{q}
  \protect \hat {Q}_{p}$ is the anharmonic Hamiltonian}\BibitemShut {NoStop}%
\bibitem [{\citenamefont {Shtogun}\ and\ \citenamefont
  {Woods}(2010)}]{Shtogun2010}%
  \BibitemOpen
  \bibfield  {author} {\bibinfo {author} {\bibfnamefont {Y.~V.}\ \bibnamefont
  {Shtogun}}\ and\ \bibinfo {author} {\bibfnamefont {L.~M.}\ \bibnamefont
  {Woods}},\ }\bibfield  {title} {\bibinfo {title} {{Many-Body van der Waals
  Interactions between Graphitic Nanostructures}},\ }\href
  {https://doi.org/10.1021/jz100309m} {\bibfield  {journal} {\bibinfo
  {journal} {J. Phys. Chem. Lett.}\ }\textbf {\bibinfo {volume} {1}},\ \bibinfo
  {pages} {1356} (\bibinfo {year} {2010})}\BibitemShut {NoStop}%
\bibitem [{\citenamefont {Silvestrelli}(2008)}]{Silvestrelli2008}%
  \BibitemOpen
  \bibfield  {author} {\bibinfo {author} {\bibfnamefont {P.~L.}\ \bibnamefont
  {Silvestrelli}},\ }\bibfield  {title} {\bibinfo {title} {{Van der Waals
  Interactions in DFT Made Easy by Wannier Functions}},\ }\href
  {https://doi.org/10.1103/PhysRevLett.100.053002} {\bibfield  {journal}
  {\bibinfo  {journal} {Phys. Rev. Lett.}\ }\textbf {\bibinfo {volume} {100}},\
  \bibinfo {pages} {053002} (\bibinfo {year} {2008})}\BibitemShut {NoStop}%
\bibitem [{\citenamefont {Tkatchenko}\ \emph {et~al.}(2013)\citenamefont
  {Tkatchenko}, \citenamefont {Ambrosetti},\ and\ \citenamefont
  {DiStasio}}]{Tkatchenko2013}%
  \BibitemOpen
  \bibfield  {author} {\bibinfo {author} {\bibfnamefont {A.}~\bibnamefont
  {Tkatchenko}}, \bibinfo {author} {\bibfnamefont {A.}~\bibnamefont
  {Ambrosetti}},\ and\ \bibinfo {author} {\bibfnamefont {R.~A.}\ \bibnamefont
  {DiStasio}},\ }\bibfield  {title} {\bibinfo {title} {{Interatomic methods for
  the dispersion energy derived from the adiabatic connection
  fluctuation-dissipation theorem}},\ }\href
  {https://doi.org/10.1063/1.4789814} {\bibfield  {journal} {\bibinfo
  {journal} {J. Chem. Phys.}\ }\textbf {\bibinfo {volume} {138}},\ \bibinfo
  {pages} {074106} (\bibinfo {year} {2013})}\BibitemShut {NoStop}%
\bibitem [{\citenamefont {Ambrosetti}\ \emph {et~al.}(2014)\citenamefont
  {Ambrosetti}, \citenamefont {Reilly}, \citenamefont {DiStasio},\ and\
  \citenamefont {Tkatchenko}}]{Tkatchenko2014}%
  \BibitemOpen
  \bibfield  {author} {\bibinfo {author} {\bibfnamefont {A.}~\bibnamefont
  {Ambrosetti}}, \bibinfo {author} {\bibfnamefont {A.~M.}\ \bibnamefont
  {Reilly}}, \bibinfo {author} {\bibfnamefont {R.~A.}\ \bibnamefont
  {DiStasio}},\ and\ \bibinfo {author} {\bibfnamefont {A.}~\bibnamefont
  {Tkatchenko}},\ }\bibfield  {title} {\bibinfo {title} {{Long-range
  correlation energy calculated from coupled atomic response functions}},\
  }\href {https://doi.org/10.1063/1.4865104} {\bibfield  {journal} {\bibinfo
  {journal} {J. Chem. Phys.}\ }\textbf {\bibinfo {volume} {140}},\ \bibinfo
  {pages} {18A508} (\bibinfo {year} {2014})}\BibitemShut {NoStop}%
\bibitem [{\citenamefont {St{\"{o}}hr}\ \emph {et~al.}(2019)\citenamefont
  {St{\"{o}}hr}, \citenamefont {Van~Voorhis},\ and\ \citenamefont
  {Tkatchenko}}]{Tkatchenko2019}%
  \BibitemOpen
  \bibfield  {author} {\bibinfo {author} {\bibfnamefont {M.}~\bibnamefont
  {St{\"{o}}hr}}, \bibinfo {author} {\bibfnamefont {T.}~\bibnamefont
  {Van~Voorhis}},\ and\ \bibinfo {author} {\bibfnamefont {A.}~\bibnamefont
  {Tkatchenko}},\ }\bibfield  {title} {\bibinfo {title} {{Theory and practice
  of modeling van der Waals interactions in electronic-structure
  calculations}},\ }\href {https://doi.org/10.1039/C9CS00060G} {\bibfield
  {journal} {\bibinfo  {journal} {Chem. Soc. Rev.}\ }\textbf {\bibinfo {volume}
  {48}},\ \bibinfo {pages} {4118} (\bibinfo {year} {2019})}\BibitemShut
  {NoStop}%
\bibitem [{\citenamefont {Ambrosetti}\ \emph {et~al.}(2022)\citenamefont
  {Ambrosetti}, \citenamefont {Umari}, \citenamefont {Silvestrelli},
  \citenamefont {Elliott},\ and\ \citenamefont {Tkatchenko}}]{Tkatchenko2022}%
  \BibitemOpen
  \bibfield  {author} {\bibinfo {author} {\bibfnamefont {A.}~\bibnamefont
  {Ambrosetti}}, \bibinfo {author} {\bibfnamefont {P.}~\bibnamefont {Umari}},
  \bibinfo {author} {\bibfnamefont {P.~L.}\ \bibnamefont {Silvestrelli}},
  \bibinfo {author} {\bibfnamefont {J.}~\bibnamefont {Elliott}},\ and\ \bibinfo
  {author} {\bibfnamefont {A.}~\bibnamefont {Tkatchenko}},\ }\bibfield  {title}
  {\bibinfo {title} {{Optical van-der-Waals forces in molecules: from
  electronic Bethe-Salpeter calculations to the many-body dispersion model}},\
  }\href {https://doi.org/10.1038/s41467-022-28461-y} {\bibfield  {journal}
  {\bibinfo  {journal} {Nat. Commun.}\ }\textbf {\bibinfo {volume} {13}},\
  \bibinfo {pages} {813} (\bibinfo {year} {2022})}\BibitemShut {NoStop}%
\bibitem [{\citenamefont {London}(1937)}]{London1937}%
  \BibitemOpen
  \bibfield  {author} {\bibinfo {author} {\bibfnamefont {F.}~\bibnamefont
  {London}},\ }\bibfield  {title} {\bibinfo {title} {{The general theory of
  molecular forces}},\ }\href {https://doi.org/10.1039/tf937330008b} {\bibfield
   {journal} {\bibinfo  {journal} {Trans. Faraday Soc.}\ }\textbf {\bibinfo
  {volume} {33}},\ \bibinfo {pages} {8b} (\bibinfo {year} {1937})}\BibitemShut
  {NoStop}%
\bibitem [{\citenamefont {Schweig}(1967{\natexlab{a}})}]{Schweig1967a}%
  \BibitemOpen
  \bibfield  {author} {\bibinfo {author} {\bibfnamefont {A.}~\bibnamefont
  {Schweig}},\ }\bibfield  {title} {\bibinfo {title} {{Calculation of static
  electric polarizabilities of closed shell organic $\pi$-electron systems
  using a variation method}},\ }\href
  {https://doi.org/10.1016/0009-2614(67)85016-4} {\bibfield  {journal}
  {\bibinfo  {journal} {Chem. Phys. Lett.}\ }\textbf {\bibinfo {volume} {1}},\
  \bibinfo {pages} {163} (\bibinfo {year} {1967}{\natexlab{a}})}\BibitemShut
  {NoStop}%
\bibitem [{\citenamefont {Schweig}(1967{\natexlab{b}})}]{Schweig1967b}%
  \BibitemOpen
  \bibfield  {author} {\bibinfo {author} {\bibfnamefont {A.}~\bibnamefont
  {Schweig}},\ }\bibfield  {title} {\bibinfo {title} {{Calculation of static
  electric higher polarizabilities of closed shell organic $\pi$-electron
  systems using a variation method}},\ }\href
  {https://doi.org/10.1016/0009-2614(67)85047-4} {\bibfield  {journal}
  {\bibinfo  {journal} {Chem. Phys. Lett.}\ }\textbf {\bibinfo {volume} {1}},\
  \bibinfo {pages} {195} (\bibinfo {year} {1967}{\natexlab{b}})}\BibitemShut
  {NoStop}%
\bibitem [{\citenamefont {Rayane}\ \emph {et~al.}(2002)\citenamefont {Rayane},
  \citenamefont {Compagnon}, \citenamefont {Antoine}, \citenamefont {Broyer},
  \citenamefont {Dugourd}, \citenamefont {Labastie}, \citenamefont
  {L’Hermite}, \citenamefont {Le~Padellec}, \citenamefont {Durand},
  \citenamefont {Calvo}, \citenamefont {Spiegelman},\ and\ \citenamefont
  {Allouche}}]{Rayane2002}%
  \BibitemOpen
  \bibfield  {author} {\bibinfo {author} {\bibfnamefont {D.}~\bibnamefont
  {Rayane}}, \bibinfo {author} {\bibfnamefont {I.}~\bibnamefont {Compagnon}},
  \bibinfo {author} {\bibfnamefont {R.}~\bibnamefont {Antoine}}, \bibinfo
  {author} {\bibfnamefont {M.}~\bibnamefont {Broyer}}, \bibinfo {author}
  {\bibfnamefont {P.}~\bibnamefont {Dugourd}}, \bibinfo {author} {\bibfnamefont
  {P.}~\bibnamefont {Labastie}}, \bibinfo {author} {\bibfnamefont {J.~M.}\
  \bibnamefont {L’Hermite}}, \bibinfo {author} {\bibfnamefont
  {A.}~\bibnamefont {Le~Padellec}}, \bibinfo {author} {\bibfnamefont
  {G.}~\bibnamefont {Durand}}, \bibinfo {author} {\bibfnamefont
  {F.}~\bibnamefont {Calvo}}, \bibinfo {author} {\bibfnamefont
  {F.}~\bibnamefont {Spiegelman}},\ and\ \bibinfo {author} {\bibfnamefont
  {A.~R.}\ \bibnamefont {Allouche}},\ }\bibfield  {title} {\bibinfo {title}
  {{Electric dipole moments and polarizabilities of single excess electron
  sodium fluoride clusters: Experiment and theory}},\ }\href
  {https://doi.org/10.1063/1.1480595} {\bibfield  {journal} {\bibinfo
  {journal} {J. Chem. Phys.}\ }\textbf {\bibinfo {volume} {116}},\ \bibinfo
  {pages} {10730} (\bibinfo {year} {2002})}\BibitemShut {NoStop}%
\bibitem [{\citenamefont {Feranchuk}\ \emph {et~al.}(2015)\citenamefont
  {Feranchuk}, \citenamefont {Ivanov}, \citenamefont {Le},\ and\ \citenamefont
  {Ulyanenkov}}]{Hoangbook2015}%
  \BibitemOpen
  \bibfield  {author} {\bibinfo {author} {\bibfnamefont {I.}~\bibnamefont
  {Feranchuk}}, \bibinfo {author} {\bibfnamefont {A.}~\bibnamefont {Ivanov}},
  \bibinfo {author} {\bibfnamefont {V.-H.}\ \bibnamefont {Le}},\ and\ \bibinfo
  {author} {\bibfnamefont {A.}~\bibnamefont {Ulyanenkov}},\ }\href
  {https://doi.org/10.1007/978-3-319-13006-4} {\emph {\bibinfo {title}
  {{Non-perturbative Description of Quantum Systems}}}}\ (\bibinfo  {publisher}
  {Springer},\ \bibinfo {address} {Switzerland},\ \bibinfo {year}
  {2015})\BibitemShut {NoStop}%
\end{thebibliography}%

\end{document}